\newif\ifanonymous
\newif\ifdraft

\draftfalse
\pdfoutput=1
\documentclass[letterpaper,twocolumn,10pt]{article}

\ifdraft
\fi

\usepackage{versions}

\excludeversion{conf}
\includeversion{full}

\usepackage{amsthm}
\usepackage{amsmath}
\usepackage{amssymb}
\usepackage{stmaryrd}
\usepackage{tabularx}
\usepackage[utf8]{inputenc}
\usepackage[table,xcdraw,rgb]{xcolor}
\usepackage{xr-hyper}
\usepackage[a-2b]{pdfx}
\usepackage{xspace}
\usepackage{proof}
\usepackage{multirow}
\usepackage{paralist}
\usepackage[inline]{enumitem}
\usepackage{thm-restate}
\usepackage{booktabs} %
\usepackage{tabularx}
\usepackage{tikz}
\usepackage{pgfplots}
\pgfplotsset{compat=1.17}
\usepackage{todonotes}
 \usepackage{booktabs}
\usetikzlibrary{backgrounds,positioning,fit,calc}
\usepgflibrary{shapes.geometric}
\colorlet{rose}{red!20}
\colorlet{lima}{yellow!30}
\colorlet{light}{black!10}
\usetikzlibrary{fit}
\usetikzlibrary{tikzmark}
\usetikzlibrary{backgrounds,positioning,fit,calc}
\usepgflibrary{shapes.geometric}
\usetikzlibrary{decorations.pathreplacing,calc}
\usetikzlibrary{arrows,shapes,positioning,shadows,trees,automata,tikzmark}
\usepackage{booktabs}
\usepackage[binary-units=true]{siunitx}
\usepackage{msc}
\usepackage{subcaption}
\usepackage{calc}

\usepackage{float}
\usepackage{listings} %

\usepackage{hyperref}

\definecolor{cyanCYMK}{RGB}{0,163,218}
\hypersetup{
  filecolor       = {cyanCYMK},
}

\usepackage[capitalize]{cleveref} %
\usepackage{mathtools}

\usepackage{usenix-2020-09}
\usepackage{DejaVuSansMono}

\makeatletter
\renewcommand\paragraph{\@startsection{paragraph}{4}{\z@}%
                                   {2ex \@plus1ex \@minus.2ex}%
                                   {-1em}%
                                   {\normalfont\normalsize\bfseries}}
\makeatother

 \usepackage[
 	backend=biber,
    style=numeric,
    natbib=true,
    url=false,
    doi=false,
    isbn=false,
    eprint=false,
    maxbibnames=9
]{biblatex}

\usepackage{xurl}

\processifversion{conf}{\addbibresource{full.bib}}

\renewbibmacro{url+urldate}{}

\newbibmacro{string+doiurl}[1]{%
  \iffieldundef{doi}{%
    \iffieldundef{url}{%
      #1
    }{%
      \href{\thefield{url}}{#1}%
    }%
  }{%
    \href{http://dx.doi.org/\thefield{doi}}{#1}%
  }%
}

\DeclareFieldFormat[
  article,
  inbook,
  incollection,
  inproceedings,
  report,
  misc,
  software,
  online,
  thesis,
  unpublished
]{title}{\usebibmacro{string+doiurl}{#1}}

\AtEveryBibitem{%
  \clearfield{date}%
  \clearfield{urlyear}%
  \clearname{editor}%
  \clearfield{issn}%
  \clearfield{isbn}%
  \clearfield{volume}%
  \clearfield{series}%
  \clearlist{publisher}%
}

\crefname{cnd}{condition}{conditions}
\lstdefinelanguage{tamarin}{
  sensitive=true,
  morecomment=[l]{//},
  morestring=[b]',
  morestring=[s]{`}{'},
}

\lstdefinestyle{tamarin}{
    language={tamarin},
    escapechar={*},
}

\tikzset{
	basic/.style       = {draw, drop shadow, font=\small\ttfamily, rectangle},
	dom/.style         = {basic, rounded corners=2pt, thick, align=center, fill=white, draw=blue},
	ip/.style          = {basic, rounded corners=2pt, thick, align=center, fill=white, draw=green},
	org/.style         = {basic, rounded corners=2pt, thick, align=center, fill=white, draw=red},
	cc/.style          = {basic, rounded corners=2pt, thick, align=center, fill=white, draw=purple},
	as/.style          = {basic, rounded corners=2pt, thick, align=center, fill=white, draw=orange},
	label/.style       = {font=\small\sffamily, text centered},
	ex1/.style         = {basic, rounded corners=2pt, thick, align=center, fill=white, draw=black},
	prop/.style         = {basic, dashed, fill=white, draw=black}
}

\newcommand{\mathcmd}[1]{{\normalfont\ensuremath{#1}}\xspace}

\newcommand{\mathvalue}[1]{\mathcmd{\mathit{#1}}}

\newcommand{\mathlabel}[1]{\mathcmd{\textsf{#1}}}

\newcommand{\textop}[1]{\relax\ifmmode\mathop{\text{#1}}\else\text{#1}\fi}

\newcommand{\msgsort}{\ensuremath{\mathit{msg}}\xspace}

\newcommand{\tempsort}{\ensuremath{\mathit{temp}}\xspace}

\newcommand{\mathfact}[1]{\mathlabel{#1}}

\newcommand{\K}{\ensuremath{!\mathfact{K}}\xspace}

\newcommand{\filter}{\ensuremath{\mathit{filter}}\xspace}

\newcommand{\theactualrule}[1]{\text{Please redefine the command
theactualrule.}}
\newcommand{\underscorethingy}[1]{\text{Please redefine the command
underscorethingy.}}

\newcommand{\hash}{\ensuremath{\mathlabel{h}}}
\newcommand{\senc}{\ensuremath{\mathlabel{senc}}}

\newcommand{\sdec}{\ensuremath{\mathlabel{sdec}}}

\makeatletter
\DeclareRobustCommand{\defeq}{\mathrel{\rlap{%
  \raisebox{0.3ex}{$\m@th\cdot$}}%
  \raisebox{-0.3ex}{$\m@th\cdot$}}%
  =}
\DeclareRobustCommand{\eqdef}{=\mathrel{\rlap{%
  \raisebox{0.3ex}{$\m@th\cdot$}}%
  \raisebox{-0.3ex}{$\m@th\cdot$}}%
  }
\makeatother

\newtheorem{definition}{\textbf{Definition}}

\theoremstyle{remark}

\newcommand{\ledot}{\mathrel{\ooalign{\hss\raise.200ex\hbox{$\cdot$}\hss\cr$\le$}}}
\newcommand{\gedot}{\mathrel{\ooalign{\hss\raise.200ex\hbox{$\cdot$}\hss\cr$\ge$}}}

\newtheoremstyle{component}{}{}{}{}{\itshape}{.}{.5em}{\thmnote{#3}#1}
\theoremstyle{component}

\newcommand{\R}{\ensuremath{\mathbb R}}

\newcommand{\dptt}{DP3T}
\newcommand{\robert}{ROBERT}
\newcommand{\cwa}{CWA}
\newcommand{\gaen}{GAEN}

\newcommand{\ts}{\mathvalue{ts}}

\newcommand{\LEE}{\mathlabel{LEE}}

\newcommand{\id}{\mathvalue{id}}
\newcommand{\epoch}{i}

\newcommand{\SymKeyServer}{K_S}
\newcommand{\PubKeyServer}{pk_S}
\newcommand{\PriKeyServer}{sk_S}
\newcommand{\FedKeyServer}{K_\mathit{fed}}
\newcommand{\CountryCode}{CC_S}
\newcommand{\PubKeyApp}{pk_A}
\newcommand{\PriKeyApp}{sk_A}
\newcommand{\EncKeyApp}{K_A^\mathit{Enc}}
\newcommand{\AuthKeyApp}{K_A^\mathit{Auth}}
\newcommand{\SharedSecretServerApp}{K_\mathit{SA}}

\newcommand{\EBID}[2]{\mathit{EBID}_{{#1},{#2}}}
\newcommand{\ECC}[2]{\mathit{ECC}_{{#1},{#2}}}
\newcommand{\sencBra}[2]{\{{#2}\}_{#1}}
\newcommand{\MAC}[2]{\mathit{HMAC}_{#1}(#2)}

\newcommand{\tfact}[1]{\mathlabel{#1}}
\newcommand{\SpaceTime}[1]{\tfact{!SpaceTime}}
\newcommand{\Time}[1]{\tfact{SpaceTime}}

\newcommand{\PClaimAtRisk}{\tfact{PClaimAtRisk}}
\newcommand{\HAClaimInfected}{\tfact{HAClaimInfected}}

\newcommand{\WithinFTDays}{\tfact{Within14Days}}
\newcommand{\IsAt}{\tfact{IsAt}}

\newcommand{\place}{\mathvalue{p}}

\newcommand{\qr}{\mathvalue{qr}}
\newcommand{\eph}{\mathvalue{eph}}
\newcommand{\guid}{\mathvalue{guid}}
\newcommand{\rg}{\mathvalue{regToken}}
\newcommand{\cwatan}{\mathvalue{tan}}
\newcommand{\tek}{\mathvalue{tek}}
\newcommand{\AC}{\mathvalue{AC}}
\newcommand{\cc}{\mathvalue{cc}}

\newcommand{\TEK}[1]{\mathit{TEK}_{{#1}}}
\newcommand{\HKDF}[1]{\mathit{HKDF}({#1})}
\newcommand{\RPIK}[1]{\mathit{RPIK}_{#1}}
\newcommand{\AEMK}[1]{\mathit{AEMK}_{#1}}
\newcommand{\RPI}[2]{\mathit{RPI}_{{#1},{#2}}}
\newcommand{\AEM}[2]{\mathit{AEM}_{{#1},{#2}}}

\newcommand{\anycc}{$^\mathit{any}$\xspace}
\newcommand{\samecc}{$^\mathit{same}$\xspace}

\newcommand{\dayf}{$^\mathit{day}$\xspace}
\newcommand{\obtainf}{$^{*}$\xspace}
\newcommand{\maxuploadf}{\xspace}

\newcommand{\outsidef}{$^{**}$\xspace}

\newcommand{\li}[1]{\lstinline[style=tamarin]{#1}}

\newlist{tabitemize}{itemize}{1}
\setlist[tabitemize]{label=\textbullet,nosep,before=\compress,leftmargin=*,itemsep=1pt}
\makeatletter
\newcommand*{\compress}{\@minipagetrue}
\makeatother

\definecolor{azure}{rgb}{0.03, 0.08, 1.00}

\ifdraft%
\newenvironment{nnew}{\color{azure}}{}
\else%
\newenvironment{nnew}{}{}
\fi

\newcommand{\fullversionref}{contacttracing-full}
\processifversion{conf}{%
\externaldocument[F-]{contacttracing-full}[contacttracing-full.pdf]
}

\newcommand{\xrref}[1]{%
\processifversion{conf}{\ref{F-#1}}%
\processifversion{full}{\ref{#1}}%
}

\newcommand{\xrcite}[1]{%
\processifversion{conf}{\cite[\Cref{F-#1}]{\fullversionref}}%
\processifversion{full}{\cref{#1}}%
}

\newcommand{\appendixorfull}[1]{%
\processifversion{conf}{the full version~\cite[\Cref{F-#1}]{\fullversionref}}%
\processifversion{full}{\Cref{#1}}%
}

\begin{document}

\ifanonymous \else
  \author{
    {\rm Kevin Morio$^1$ $\quad$ Ilkan Esiyok$^1$ $\quad$ Dennis Jackson$^2$ $\quad$ Robert Künnemann$^1$}\\\\
    $^1$CISPA Helmholtz Center for Information Security \\ 
    $^2$Mozilla
  }
\fi

\title{%
    Automated Security Analysis of Exposure Notification Systems
    \processifversion{full}{\\(Full Version)\thanks{This is an extended version of \cite{morio2023automated}.}}
}

\maketitle
\processifversion{full}{\enlargethispage{.25\baselineskip}}

\begin{abstract}
    We present the first formal analysis and comparison of the security of
the two most widely deployed exposure notification systems:
the ROBust and privacy-presERving proximity Tracing protocol (\robert{}) and
the Google Apple Exposure Notification (\gaen{}) framework.

\robert{} is the most popular instalment of the centralised approach to
exposure notification, in which the risk score is computed by
a central server. \gaen{}, in contrast, follows the
decentralised approach, where the user's phone calculates the risk.
The relative merits of centralised and decentralised systems have
proven to be a controversial question. The majority of the previous analyses have focused on the privacy implications of these systems, ours is the first formal analysis to evaluate the security of the deployed systems---the absence of false risk alerts.

We model the French deployment of \robert{} and the most widely deployed \gaen{} variant, Germany's
Corona-Warn-App. We isolate the precise conditions under
which these systems prevent false alerts. We determine exactly how an
adversary can subvert the system via network and Bluetooth
sniffing, database leakage or the compromise of phones,
back-end systems and health authorities. We also investigate the security of the original specification of the \dptt{} protocol, in order to identify gaps between the proposed scheme and its ultimate deployment. 

We find a total of 27 attack patterns, including
many %
that
distinguish the centralised from the decentralised approach, as well as
attacks on the authorisation procedure that differentiate all three
protocols.
Our results suggest that \robert{}'s centralised design is more vulnerable against 
both opportunistic and highly resourced attackers 
trying to perform mass-notification attacks. 

\end{abstract}

\section{Introduction}

In response to the COVID-19 pandemic, digital exposure notification systems (ENS) have been designed, deployed and are now used by hundreds of millions of people around the world. These systems complement manual contact tracing efforts, providing automated notification to users that were potentially exposed to COVID-19 allowing them to take appropriate precautions.

Although various designs have been proposed, the systems which have seen real world deployment can be split into two families: centralised~\cite{privaticsteaminriaROBERTROBustPrivacypresERving}
or
decentralised~\cite{troncosoDecentralizedPrivacyPreservingProximity2020},
depending on whether the risk calculation is performed on a central server or on the user's device.
The merits of both approaches have been the subject of extensive debate, with much of the argument focused on potential privacy risks.
Less attention, however, was paid to the comparative security of the two categories of systems.

In this work, we formally analyse the security of the two most widely
deployed centralised and decentralised ENS: the ROBust and
privacy-presERving proximity Tracing protocol (\robert{}), the leading
centralised solution, and the Google Apple Exposure Notification
framework (\gaen{}), which follows a decentralised
design.
Whilst \robert{} was proposed and developed as a cohesive whole, the \gaen{} framework leaves many security-critical components of the system open for the developers to implement.
The countries that deployed applications based on the \gaen{} framework thus had to decide
how to authenticate positive test results,
how to disseminate this information to a central back end,
and
how to interoperate with other countries’ back ends.
Clearly, these design decisions critically impact the overall security of the system.
Consequently, we focus on one of the most popular applications built with the \gaen{} framework, the
Corona-Warn-App (\cwa{}) developed in Germany, but also investigate the security
of the \dptt{} system as originally specified. \dptt{}'s specification informed the design of
\gaen{}-based applications deployed in
Austria, Belgium, Croatia, Ireland, Italy, the Netherlands, Portugal
and Switzerland.

We carry out our formal analysis with Tamarin~\cite{SMCB-cav13}, a security protocol verifier.
Our models include all the components of the deployed ENS including the user phones, a granular temporal and spatial model and the various back-end components and associated protocols.
Our models allow for an unbounded number of users engaging in an unbounded number of sessions and allow the adversary to control user location and travel patterns, COVID-19 diagnostics and network communication. We further consider an attacker who can compromise any entity in the system, including users' phones, health authorities and national back ends.
For each of the models, we ask the following questions:
\begin{enumerate}[label=\alph*)]
    \item Under what circumstances can an attacker pollute the ENS with forged information which will impact the risk computation?
    \item Under what circumstances can an attacker mislead a user into believing they are at risk of infection?
\end{enumerate}
By finding answers to these questions, we develop an ontology of possible attacks which systematises the result of much previous work claiming various attacks or proposing improvements without evidence (discussed in \cref{sec:related}).
The generality of our threat model allows us to compare the discussed systems with the required level of detail.

We discover attack patterns that are common across all systems, but also
attacks that distinguish centralised and decentralised systems.
This is fairly obvious in scenarios with back end compromise, but we
also discover an important difference in the scalability of an attack
when a phone of an infected user is under the adversary's control.
While isolated false alarms can severely impact individuals, a large-scale
attack can render the whole system unusable, severely impacting epidemic control efforts.

Independent of the architecture, the authentication schemes used in
the three proposals give vastly different guarantees, which is largely
because interactive schemes are hard to deploy at the
necessary scale.
We also find that federation increases the vulnerability to
attacks on other countries' health authorities, with a stronger impact on centralised than decentralised systems.
In total, we find 27 attack patterns across \robert{}, \dptt{} and the \cwa{}, providing a detailed answer of the two previously posed questions.

\paragraph*{Contributions}
This paper makes the following contributions:
\begin{enumerate*}[label=(\alph*)]
	\item The first formal security analysis of the most widely deployed ENS.
    \item A systematic evaluation of the trade-offs
          between these systems w.r.t.\ a powerful shared threat model.
	\item Novel techniques for modelling spatial and temporal
          aspects of these protocols. They are amenable for
          automated analysis and
          can be reused to model proposed refinements, extensions and designs.
\end{enumerate*}

\paragraph*{Outline}

In \cref{sec:related}, we discuss related work on the security of ENS.
We then introduce \robert{}, \dptt{} and the \cwa{} in \cref{sec:ens}.
The modelling of the systems is discussed in \cref{sec:formalisation} and their security properties in \cref{sec:goals}.
The results of our analysis are presented in \cref{sec:results}, limitations in \cref{sec:limitations}.
Finally, we conclude in \cref{sec:conclusion}.

\section{Related Work}\label{sec:related}

A substantial amount of previous work has proposed alternative ENS~\cite{casagrandeContactTracingMade2020,pietrzakDelayedAuthenticationPreventing2020,pinkashashomer} and presented informal assessments of their properties~\cite{reichert2021survey,dp3tDP3TUploadAuthorisation,aisecPandemicContactTracing2020,vaudenayCentralizedDecentralizedContact2020,baumgartnerMindGAPSecurity2020,austriaInverseSybilAttacksAutomated2020,avitabileTEnKUTerroristAttacks2020}.
However, only three~\cite{danzSecurityPrivacyDecentralized2020,kobeissiVerifpalCryptographicProtocol2019,canettiPrivacyPreservingAutomatedExposure2020} undertook a formal analysis of the security properties of ENS.
In this section, we review these in detail, postponing a discussion of previously published attacks until we discuss our taxonomy in \cref{sec:results}.

\textcite{danzSecurityPrivacyDecentralized2020} investigate the security and privacy of the Temporary Contact Numbers Protocol and the early \dptt{} proposals, as well as the currently deployed \gaen{} framework.
Their analysis is focused on the `cryptographic core' of each proposal, specifically the Bluetooth proximity protocol.
Unlike our work, they do not consider the security of the authorisation procedure for uploads and do not model the back end infrastructure or the spatial distribution of users.
However, as they carry out their analysis by hand in the computational model, they are able to define and prove various privacy properties for the respective Bluetooth proximity protocols.

\textcite{kobeissiVerifpalCryptographicProtocol2019} build a model of \dptt{} in order to demonstrate the flexibility and friendliness of their new protocol analysis tool.
As the model was developed as an example, it only captures some parts of the protocol, limited to only three participants interacting in a prescribed pattern without an authorisation procedure.

\textcite{canettiPrivacyPreservingAutomatedExposure2020} propose two novel ENS and investigate their security and privacy.
They formalise their security notions partly in the UC (universal composability) model and partly via game based definitions.
Although their UC model provides a computational proof that the modelled systems meet their specified ideal functionality, the concrete implications of this ideal functionality are not explored and extracting the resulting security properties is a considerable challenge.
Similar to \textcite{danzSecurityPrivacyDecentralized2020}, they model their proposed Bluetooth proximity protocols in detail, but leave the rest of their system, e.g. back end infrastructure and upload authorisation, abstract.

\section{Exposure Notification Systems}\label{sec:ens}

\begin{nnew}In this section, we describe the design of \robert{}, \gaen{}, \dptt{} and the \cwa{}.

We first introduce \robert{}, the centralised ENS designed by INRIA PRIVATICS and Fraunhofer AISEC~\cite{privaticsteaminriaROBERTROBustPrivacypresERving}.
\robert{} was proposed in April 2020
and deployed in the official French contact tracing application `StopCovid' on 2 June 2020 (later rebranded to `TousAntiCovid').
With over 12.3 million downloads, it is the most popular centralised ENS.

We then describe the \gaen{} framework which underlies many deployed decentralised systems.
It is a library integrated with Google's Android OS and Apple's iOS to provide a Bluetooth proximity protocol.
Next, we detail \dptt{} which predates the \gaen{} specification and is substantially similar in design, 
but also describes the various other components of ENS
and guides the design of many European exposure notification applications. 
We also discuss the \dptt{} proposal for user upload authentication in detail.
Finally, we elaborate on the design of the \cwa{}, the official German exposure notification application, based on \dptt{} and the \gaen{} framework.
\end{nnew}

\paragraph{Terminology}%
As we refer to the phone more frequently than to its owner, 
we will call a phone \emph{infected} or \emph{diagnosed} if its owner
has been infected (or diagnosed) with COVID-19.
The \emph{contagious period} is the time period in which the phone's owner is assumed to be contagious.
Ephemeral keys that are
associated with this period
are called \emph{infected keys}.
If two phones are close enough that one's beacon message
reaches the other, they are \emph{in proximity}.

\begin{figure*}%
	\centering
        \noindent\begin{minipage}{\textwidth}
        \begin{minipage}[c][6cm][c]{.5\textwidth}
        \includegraphics[width=\linewidth]{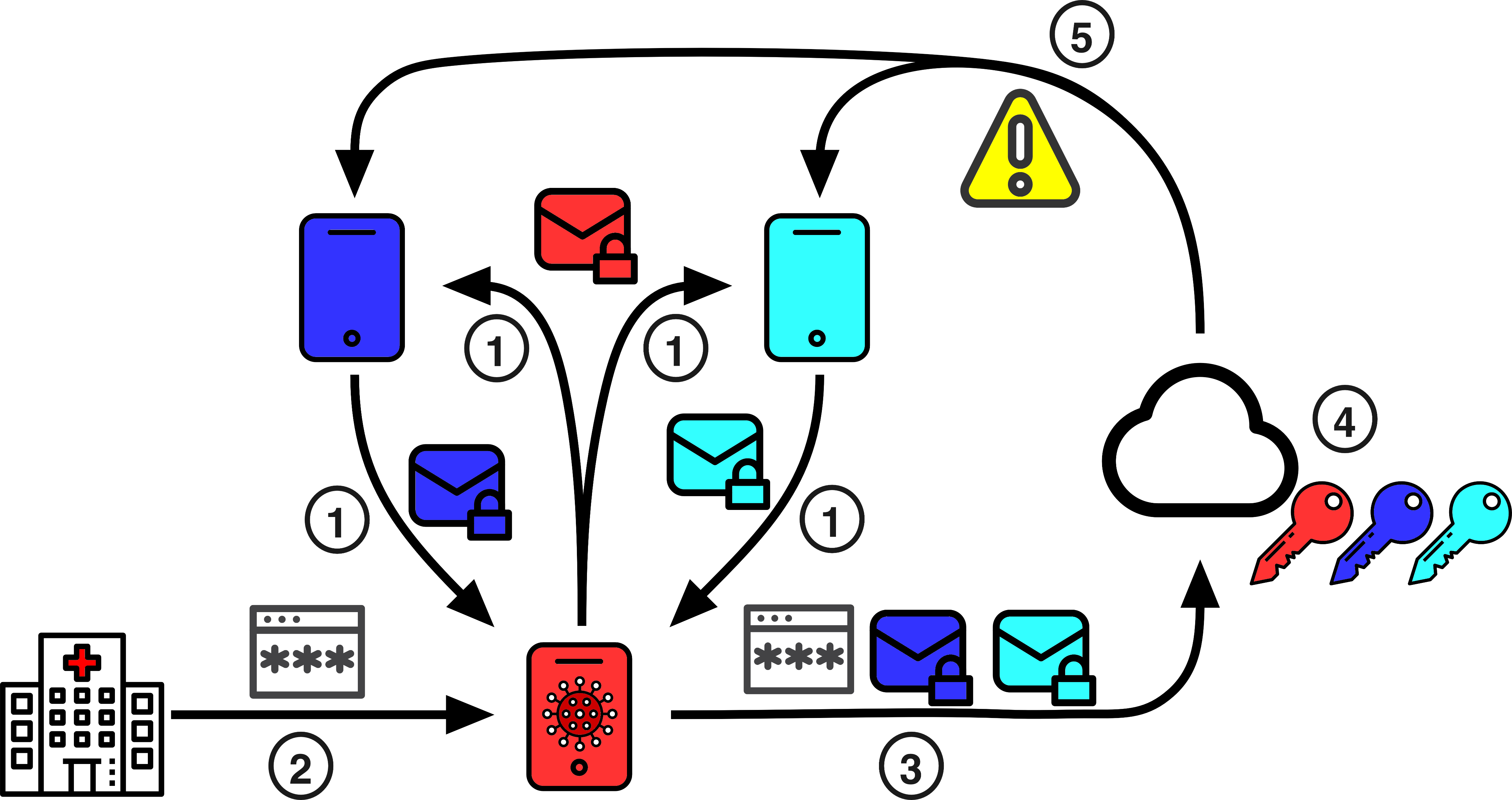}
        \end{minipage}\hfill
        \begin{minipage}[c][6cm][c]{.5\textwidth}
	\includegraphics[width=\linewidth]{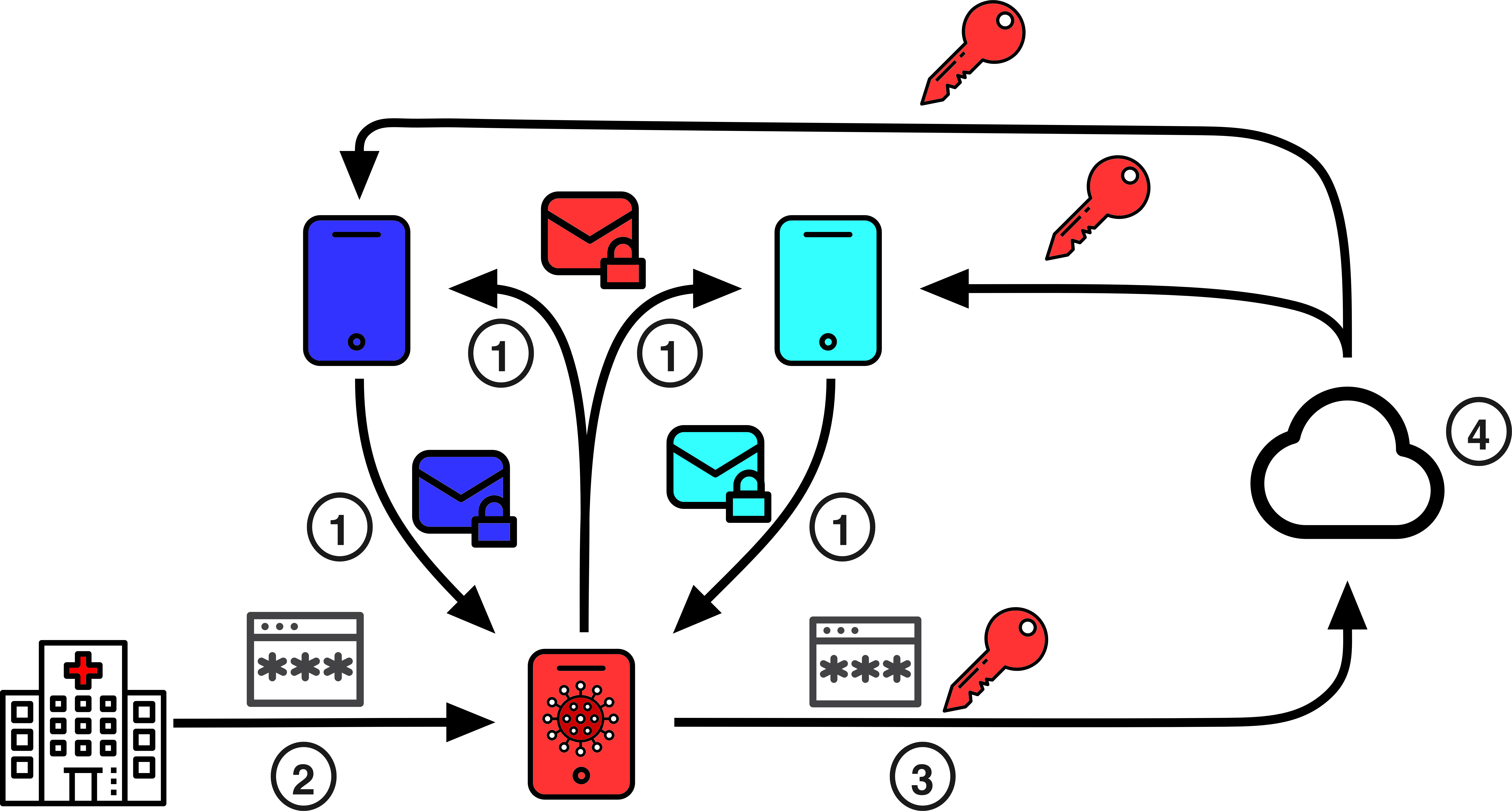}
        \end{minipage}%
        \end{minipage}
        \caption{An overview of \robert{} (left) and the \gaen{} framework (right). The colour of the icons on an edge refers to the entity which originally generated that value.
	\label{fig:gaen-system}
    \label{fig:an-overview-over-the-robert-system-}}
\end{figure*} %
\subsection{\robert{}}

In \robert{},
mobile phones continuously broadcast ephemeral identifiers and store the identifiers they receive from other users.
If a user is diagnosed with COVID-19, they upload all the recorded ephemeral identifiers on their phone to a central back end, which is then able to link these ephemeral identifiers to the users that broadcast them, carry out a risk calculation and notify any impacted users.

For linking ephemeral identifiers to users, phones must first register with a central server, which assigns them a permanent identifier, gives them an authorisation key and stores an encryption key for them.
The server then periodically generates batches of ephemeral Bluetooth identifiers (EBIDs) and transmits them to the phone.

In normal operation (\cref{fig:an-overview-over-the-robert-system-}), phones continuously broadcast their EBID for the current epoch and record any received EBIDs transmitted by other phones (1).
Should a user test positive, the health authority provides them with a special authorisation code which they enter into their phone (2).
The phone then transmits this code and all witnessed EBIDs during the contagious period to the central server (3).
The server verifies the authorisation code with the help of the health authority (not shown).
Knowing each user's encryption key, the server can decrypt the EBIDs and learn the permanent identifiers of each user who has been in proximity to the contagious user (4).
After performing a risk calculation, the central server informs the impacted users of the risk (5), but does not disclose any information about the cause of the risk.

\begin{nnew}
The specification explicitly leaves the authorisation procedure for step (2) open, but requires that `only authorised and positive-tested users are allowed to upload'.
We thus investigated the source code of the back-end server, submission code server and the REST API specification%
\footnote{Available under \url{https://gitlab.inria.fr/stopcovid19}.
    Note that we cannot be sure
    \begin{enumerate*}[label=(\alph*)]
        \item that this code actually runs on the server and
        \item if system parameters like token validity deviate from
            their
            defaults.
    \end{enumerate*}
    \label{foot:sourcecode}} %
and found that the submission code server produces \emph{long codes} for health professionals with a default expiration time of 8 days and \emph{short codes} 
with a default expiration time of 60 minutes.
A positively tested user is given such a code in form of a QR code, which can be scanned to initiate the upload of EBIDs.\end{nnew}

\robert{} also supports a federated deployment, in which different \robert{} deployments can interoperate with one another.
In this case, all servers in the federation are provided with a federation key which is used to encrypt a country code for each EBID.
When receiving EBIDs from an infectious user, the back end can always learn the country code of an EBID and forward the EBID to the relevant server for processing.
We provide a more detailed description of registration, broadcast, authorisation, risk calculation and federation in \cref{sec:robert-details}.

\subsection{\gaen{}}

In April 2020, Google and Apple announced a framework for
decentralised ENS (substantially similar to previous proposals
such as \dptt{}, PACT\footnote{\href{https://pact.mit.edu}{Privacy Automated Contact Tracing}} and TCN\footnote{\href{https://github.com/TCNCoalition/TCN}{Temporary Contact Numbers}}) which they deployed via their respective mobile phone operating systems.
This framework implements the broadcast protocol and risk calculation,
but leaves the other components such as the interaction with the
health authority and between back ends to the developers of the
country-specific
contact tracing applications, such as the \cwa{} and other \dptt{} implementations.
In the \gaen{} framework (\cref{fig:gaen-system}), phones do not need to register with any central provider.
Instead they generate a fresh secret key at the start of each day, called the Temporary Exposure Key (TEK), which is then used to generate ephemeral identifiers.
Those are broadcast (1) and recorded by other nearby phones.
If a user is diagnosed as infected, they will receive authorisation from a health authority (2) to upload their tracing keys to the system's back end (3).
The back end validates the authorisation token and stores the keys (4).
These can then be downloaded by other users of the system (5), who can use the information to evaluate whether they have been near the infected phone and if necessary, notify the user.

Unlike \robert{}, infected phones upload their own keys instead of payloads received from other phones.
Further, the central server does not process user data and instead acts as a bulletin board for contagious keys which are distributed to all users.
We provide a more detailed description of broadcast and risk calculation in \cref{sec:robert-details}.

\subsection{\dptt{}}

Prior to the release of the \gaen{} framework, the \dptt{} ENS was proposed by a pan-European group of researchers~\cite{troncosoDecentralizedPrivacyPreservingProximity2020} in March 2020.
\dptt{} uses a Bluetooth protocol very similar to the \gaen{} framework, but, in contrast, also proposed a number of authorisation procedures.

In the \gaen{} framework, users who have been diagnosed with COVID-19 must upload their TEKs to a central server, which distributes them to all other users to begin the risk evaluation process.
Consequently, the security of this upload procedure is critical to the overall system, as users could maliciously upload their TEKs to cause a false alert and trigger quarantine procedures for users who are not actually at risk.

In the vast majority of \gaen{} deployments, the user must be diagnosed as infected with COVID-19 by a health professional.
The \dptt{} consortium proposed three different authorisation protocols, which detail how the user can authenticate with a health authority and use the resulting credential to upload their TEKs to the back end.
Variants 1 and 2 are very similar to the ones used by the \cwa{} which we describe in \cref{sec:cwa}.
We describe Variant 3, designed to have the strongest security properties, below.

\paragraph{Device Bound Authorisation Codes}

\begin{figure}[ht!]
    \centering
    \input{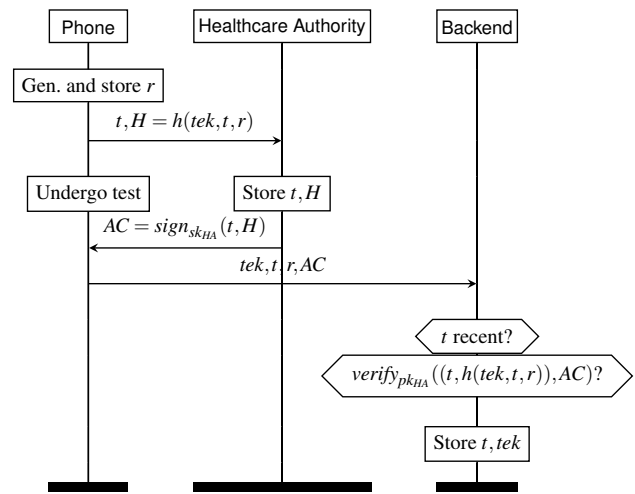}
    \caption{Authorisation procedure for Variant 3 of \dptt{}. Let $t$ be the starting epoch number of the $\tek$ and $r$ a random value. For each $\tek$ in the contagious period, an additional $H$ and $\AC$ are exchanged, up to a maximum of 14.}
    \label{msc:dp3t-auth-3}
\end{figure}

The three-party protocol between the user, health authority and back end is depicted in \cref{msc:dp3t-auth-3}.
The user generates a blinded commitment of their previous TEKs and transmits them (via the Internet or by displaying a QR code) to the health authority before being tested.
Later, if the test is positive, the health authority provides the user with an authorisation code (a digital signature).
The user can then transmit this authorisation code to the back end which checks if the code is both valid and recent.
The user is assumed to share an authentic channel with the health authority and, separately, with the back end.

This protocol is claimed to ensure that only users who have tested positive can upload and that the keys they upload correspond to those on the device at testing time.
Hence, any user wishing to upload malicious values (e.g. from another user) must tamper with their device before being tested.
This is argued to significantly raise the cost of any attack on the system, as an attacker must coerce (persuade, bribe) individuals prior to their testing, rather than only targeting those who test positive.

\begin{nnew}
\paragraph{Federation}
In the \gaen{} framework, each country must provide its own back end and national infrastructure.
This raises the question of how to handle travellers between countries, who are typically a high-risk category for COVID-19 exposure.
In \dptt{}, phones inform their `home` back end about regions they frequently visit, which then forwards tracing data from these regions’ back ends to the phone.\end{nnew}

\subsection{\cwa{}}\label{sec:cwa}

The Corona-Warn-App was developed by SAP SE and Deutsche Telekom and is based on the \gaen{} framework.
It uses a different authorisation scheme from \dptt{}.
Users are issued a QR code when they visit a testing facility.
They scan the code with their device, which automatically retrieves their test result.
If the test is positive, they can request a TAN (transaction authentication number) that can be used once to upload the TEKs.
As this process is vulnerable to human error, e.g. forgetting to scan the QR code or the medical facility not providing the code, a telephone authentication option (teleTAN) is offered as backup.
These two authentication protocols are similar to \dptt{} variants 1 and 2.
They are detailed further in \cref{sec:cwa-details}.

\begin{nnew}
\paragraph{European Federation Gateway Service}
The European Federation Gateway Service~\cite{ehealthnetworkEuropeanProximityTracing} was developed to provide a solution across the EU.
Each country's back end connects to a central database server via mutually authenticated TLS.
See \cref{sec:cwa-details} for more details.\end{nnew}

\section{Formalisation}\label{sec:formalisation}

\begin{nnew}
In this section, we discuss the scope of our formal models and the decisions we had to make.
We have created models for \robert{}, \dptt{} and the \cwa{} (using the \gaen{} framework) in Tamarin.%
\footnote{The models are available in the \href{https://github.com/tamarin-prover/tamarin-prover/tree/develop/examples/usenix23-ens}{Tamarin repository}.}
In cases where the specifications lack important details or are ambiguous, we examined the public source code of \robert{} and the \cwa{} to inform our decisions.
Limitations of our approach are discussed in more detail in \cref{sec:limitations}.
\end{nnew}
\paragraph{Tamarin}

We perform our analysis with Tamarin~\cite{SMCB-cav13}, a protocol verification tool in the Dolev-Yao model.
In this model, messages are described as abstract terms composed from function symbols that represent cryptographic primitives.
The behaviour of these cryptographic primitives is specified by an equational theory on terms.

For example, a hash function is represented by function symbol $\hash$ and the empty equational theory, essentially describing it as a random oracle.
Symmetric encryption and decryption are typically written as the function symbols $\senc$, $\sdec$ and the equation $\sdec(k,\senc(k,m)) = m$, which describes the behaviour of the decryption function.
A term is either an atomic value, a variable or composed from other
terms with a function symbol.
Atomic values are either public names, which are known to the
attacker and protocol, or fresh names, which are drawn uniquely and
model keys and nonces.

Protocols are modelled as a combination of multiset rewrite rules (MSRs) and trace-based `restrictions'.
MSRs describe the possible actions in a protocol, such as sending a particular network message or updating local state.

\begin{nnew}
Consider the following example:
\begin{lstlisting}
[ Phone(id,t,k) ] --[ UseKey(id,k) ]-> [ Out(senc(k,t)) ]
\end{lstlisting}
$\tfact{Phone}(\ldots)$, $\tfact{UseKey}(\ldots)$ and $\tfact{Out}(\ldots)$ are \emph{facts} containing one or more \emph{terms} (like $\senc(k,t)$).
The state is modelled as a multiset of facts and is rewritten by subsequent application of MSRs.
An MSR can be applied if the facts on the left-hand side ($\tfact{Phone}(\ldots$)) are part of the state.
When the MSR is applied, these facts are removed and substituted by the facts on the right-hand side ($\tfact{Out}(\ldots)$).
The transition is labelled with one or more events, also called action facts, in the middle of the rule ($\tfact{UseKey}(\ldots)$).

The trace of a system is the sequence of events starting from an empty state.
A set of default rules defines the adversary behaviour,
incorporating the special facts $\tfact{Out}$ and $\tfact{In}$ for
protocol output and input.

Security properties and the aforementioned restrictions are defined in the form of trace properties.
These are specified in a first-order logic with two sorts, $\tempsort$ for time points, i.e. the position within a trace, and $\msgsort$ for terms.
The atoms of this first-order logic are: $\bot$ for false, $t_1 \approx t_2$ for term equality, $i \lessdot j$ for time point ordering, $i \doteq j$ for time point equality and $F@i$, where $F$ is an event and $i$ is a variable of sort $\tempsort$, for the appearance of $F$ at index $i$.
We use $i \neq j$ as a shorthand for $i \doteq j \implies \bot$.
Restrictions (also specified via trace properties) allow for complex conditions to be imposed on the protocol execution, e.g. checking the validity of a signature or timestamp.
A restriction is assumed to be true, while a security property is to
be verified.\end{nnew}

\paragraph{Scope}

Our models cover the entire ENS, from user devices and their interactions over Bluetooth, to their communication with the back-end servers and health authorities.
There is no limit to the number of users in the system or the sessions they engage in. 
Similarly, there may be multiple back-end systems in operation, simultaneously and alongside their associated health authorities.
The back-end systems may be in federation with one another.
By lines of code, our models are (individually) among the largest models compared to Tamarin's model repository~(see
\cref{sec:modelling-challenges}).

\paragraph{Temporal Model}\label{sec:time point-modelling}

Both \robert{} and \gaen{} divide time into two different size windows,
typically on the order of 24 hours and 10 minutes. We dub the former
`days' and the latter `epochs' for clarity. Any particular epoch
is said to belong to a unique day.
In \gaen{}, epochs are aligned with Unix time (i.e.
seconds since 1 January 1970).
\robert{}, however, aligns epochs with the starting time of each
country's back-end services, which  we model faithfully.

\begin{nnew}
The temporal model is critical for our analysis and required several
iterations to achieve the necessary performance. All three protocols
have timestamps that occur in messages, requiring a distinction
between timestamps and the time point they refer to.
For example, a particular timestamp (`Monday') can refer to many time points
during some period.

We tried and subsequently abandoned two approaches from the
literature.
The first
models timestamps as integers, which can be represented in Tamarin via
its associative and commutative operator `+`.
Reasoning about equivalence is costly in this
case. The second models a global clock which emits fresh values
for timestamps as a protocol party.
In both approaches, any message
containing a timestamp could potentially leak information via that
timestamp, as it is a~priori a secret that could be used elsewhere,
e.g. for encryption.
Hence, Tamarin's backward analysis needs to perform a case
distinction over the origin of that timestamp whenever considering
message deduction.

Instead,
we model timestamps as public names, avoiding
costly deduction because public values are
trivially known in Tamarin.
Whenever
a timestamp $e$ is used in a rule (e.g. as the current epoch), the
rule is annotated with an event (e.g. $\mathit{Epoch}(e)$), but the
choice of $e$ is arbitrary. Instead, we use restrictions to give events
a meaning, e.g. for each two rule instances with
$\mathit{Epoch}(e)$ for the same $e$, any third $\mathit{Epoch}(e')$
between those must have $e=e'$.
Other axioms relate epochs to days or require that an order on
timestamps implies an order on all corresponding time points.
An interesting observation is that we do not need a complete
characterisation of time; the protocols we consider require, e.g.
non-repetition, but not actual monotonicity. We thus defined a set of
sound but not complete axioms, some of which we will now elaborate on.
We annotated our model's MSRs with the following events.
\begin{description}
\item[$\PClaimAtRisk(\id_R,\ts)$:]
    Phone $\id_R$ claims user was at risk at timestamp $\ts$.
\item[$\tfact{Day}(\ts)$, $\tfact{Epoch}(\ts)$:]
    Whatever action is performed in this rule occurs on day/epoch $ts$
\item[$\IsAt(\id,\place,\ts)$:]
    Phone $\id$ is at place $\place$ at timestamp $\ts$.
\item[$\HAClaimInfected(\id_I,\ts_b,\ts_e)$:]
    Health authority claims 
    that the contagious period of
    the phone $\id_I$ 
    is
    $(\ts_b,\ts_e)$.
\item[$\WithinFTDays(\ts,\ts')$:]
    Timestamps $\ts$ and $\ts'$ are 14 days apart.
\end{description}

At various points in each protocol, agents may check the ordering or
distance between particular epochs and days.
We define the following events that occur whenever a rule requires
this check to be successful.
\begin{description}
  \item[$\tfact{EarlierDay}(\ts,\ts')$, $\tfact{EarlierDayEq}(\ts,\ts')$:]
      Timestamp for day $\ts$ is earlier (or equal to) $\ts'$.
  \item[$\tfact{EarlierEpoch}(\ts,\ts')$, $\tfact{EarlierEpochEq}(\ts,\ts')$:]
      Timestamp for epoch $\ts$ is earlier (or equal to) $\ts'$.
\end{description}

These define the relation that timestamps impose on the protocol run.
Their semantics are encoded via restrictions.
Consider the following instance for \dptt{},
where days are the measure of time.
\begin{lstlisting}
EarlierDayEq(ts1, ts2) <=>
  (All #t1 #t2. Day(ts1)@t1 & Day(ts2)@t2
       ==> ts1 = ts2 | (not (ts1 = ts2) & #t1 < #t2))
\end{lstlisting}
Whenever a timestamp is considered smaller than another somewhere else
in the protocol, all rule instances need to be consistent with that
order. Any action effectuated at the earlier timestamp must indeed
happen before the latter action unless they are the same.
\begin{nnew}
The case distinction in the consequent of this restriction governs the
structure of the proof. By explicitly negating the first
case in the second disjunct, we prune unnecessary proof steps.\end{nnew}

We, furthermore, require timestamps to never repeat, or more precisely,
whenever they do repeat (i.e. two actions are effectuated on the same
day) all actions in between must be assigned the same timestamp.

\begin{lstlisting}
restriction timeDay:
"All d #t1 #t3. Day(d)@t1 & Day(d)@t3 & #t1 < #t3 
     ==> (All o #t2. Day(o)@t2 & #t1 < #t2 
                               & #t2 < #t3 ==> o = d)"
\end{lstlisting}

We also relate smaller units of time to larger units of time.
For example, in \dptt{} this concerns intervals and days:

\begin{lstlisting}
restriction timeEpochInDates:
"All e d o #t1 #t3. Interval(e)@t1 & Day(d)@t1 
                  & Interval(e)@t3 & Day(o)@t3 ==> d = o"
\end{lstlisting}

\end{nnew}
\paragraph{Spatial Model}

In order to faithfully model the proximity protocol, we employ
a granular spatial model in which locations have a unique tag.

\begin{nnew}
The following rules define writing and reading a message \li{m} on the Bluetooth
channel for the location \lstinline{$place} at a day \lstinline{$d} and epoch \lstinline{$e}. Note
that the \lstinline{$} indicates a variable of sort public
(a subsort of $\msgsort$) meaning that the rule can be instantiated with arbitrary
public names for these variables.
The \lstinline{!} indicates a permanent fact that once added persists and is thus not removed by other rules.
\begin{lstlisting}
rule BLEwr:
    [ In(m) ]
  --[ Day($d), Epoch($e), BLEwr($d, $e, $place, m) ]->
    [ !SpaceTime($place, $d, $e, m) ]

rule BLErd:
    [ !SpaceTime($place, $d, $e, m) ]
  --[ Day($d), Epoch($e), BLErd($d, $e, $place, m) ]->
    [ Out(m) ]
\end{lstlisting}

The attacker can read and
write from an arbitrary location at any time, but we record the
$\tfact{BLEwr}$ or
$\tfact{BLErd}$ event when they do.
The honest protocol rules directly read from or write to
\lstinline{!SpaceTime}, but each rule that does is annotated with an
event $\tfact{IsAt}$ collecting the phone identifier, place, day and
epoch.
Two users \lstinline{id,id'} are said to occupy the
same location at the same time if they both visit the location during
the same epoch, i.e. if
\begin{lstlisting}
Ex p d e #t1 #t2. IsAt(id,  p, d, e)@t1
                & IsAt(id', p, d, e)@t2
\end{lstlisting}

\paragraph{Soundness of the Spatial Model}
In our spatial model, two users are in proximity to each other if they share the same location at the same point in time.
Users in a particular location share a local Bluetooth channel.
In this sense, locations can be thought as being overlapping circles whose size is determined by BLE's transmission radius.
Users can also travel between locations in which case the timing and direction of their movement is controlled by the adversary.

As the location can be freely chosen, a phone can be at different
locations at the same time (day or epoch). This means that proximity is not
necessarily transitive, e.g. a phone $P_1$ can be in proximity to
$P_2$ but not to $P_3$, even if $P_3$ is in proximity to $P_2$.
As Bluetooth reception is far from robust, messages transmitted may
not be received by other users at the same location. This is reflected
since reception is controlled by the adversary.
Note that proximity does not necessarily mean the users were close
enough to transmit COVID-19, just close enough to potentially transmit
Bluetooth packets.\end{nnew}

\paragraph{Cryptographic Primitives}

In \robert{}, we model the Diffie-Hellman key exchange between the user's application and the back end with a prime order group.

For the symmetric encryption used in the ECC and EBID, we employ the standard primitives and the nonce-based encryption described in~\cite{pkcs11detenc}.
The MAC over the ciphertexts is represented using the standard symbolic authentication primitive.
These are the most accurate symbolic models published for these types of primitives; however, this nonetheless entails some loss of accuracy.
For example, as in \robert{} the MAC is truncated to 40 bits, the probability of successful forgery is not negligible, yet this cannot happen in the symbolic model.

In the \cwa{}, we model key derivation using the standard primitives for HKDF.
We treat the use of AES to unroll additional keys similarly.
We do not model the Associated Encrypted Metadata, as it only contains auxiliary information for the risk calculation and is not authenticated.

\section{Security Properties and Threat Model}
\label{sec:threatmodel}
\label{sec:goals}

Intuitively, the security desired of an ENS is easy to state:
\begin{quote}
    \textit{
    The ENS should notify the user of risk if, and only if, the user was
in proximity to an individual who was diagnosed by a health authority as being contagious at that time.}
\end{quote}

However, this property becomes considerably more complex
when we consider
\begin{enumerate*}[label=(\alph*)]
    \item what `in proximity' and `at the time' means w.r.t.\ the technical constraints within which the protocols have to operate,
    \item how users and health authorities relate across countries, and
    \item when we consider the impact of the compromise of devices, communication channels or infrastructure.
\end{enumerate*}
In this section, we introduce and justify
our security definitions as well as describe our
systematisation of the attacker's capabilities.

Recall that
we model the proximity of two users by providing each
location with a unique identifier.
We treat two users at the same location during the same epoch as being
in proximity for the purposes of exposure notification.
In practice, the ENS would take into account the precise duration of
the exposure and other information, e.g. signal strength.
But these are not relevant for our evaluation of the overall design of
each system.

In each system, we have multiple designated health authorities (one
per country) who can perform diagnostic tests on users.
We associate each user with their phone, as we can only measure if
phones are in proximity, not their users.
Putting this together, we can formally state our soundness property:
\begin{definition}[Soundness]\label{def:soundness-} %
	Whenever an honest phone $\id_R$ reports to a user that they
        were in contact with an infected phone
	at time point $t_c$, then there is another phone $\id_I$ for which the
	following conditions are met
        \begin{enumerate*}[label=\emph{\alph*})]
            \item the health authority (HA) determined that $\id_I$ was contagious in the time interval $[t_b,t_e]$, \label[cnd]{it:ha}
    		\item $t_b$ and $t_e$ are no more than 14 days apart, \label[cnd]{it:diff}
	    	\item both were in proximity at time $t_c$, \label[cnd]{it:prox}
		    \item $t_c\in  [t_b,t_e]$, and \label[cnd]{it:interval}
		    \item $\id_I \neq \id_R$. \label[cnd]{it:ident}
    	\end{enumerate*}
\end{definition}%

\Cref{it:ha} expresses the integrity of the alarm.  It
requires a discretisation of time that is present in all three
protocols.
\Cref{it:diff} ensures that the health authority's time interval remains in
reasonable bounds. This is important in case of a compromised health authority.
\Cref{it:prox} will be expressed by discretising space
similarly to time. In contrast to the discretisation of time, this is
not due to the protocol, but a modelling choice. We say that two
parties are in proximity, if they sent or received a message on
the Bluetooth channel at time $t$ and place $p$, where $t$ and $p$ are
discrete values.

\Cref{it:interval} links the time of proximity to the health
authority's diagnoses. In \dptt{}, the health
authority only determines the day of the (positive) test.
Consequently, the condition is weakened in this case.
\Cref{it:ident} expresses that alarms cannot be caused by
reflection attacks. Such an attack is not a big problem in itself---a risk
notification to a user who has already been diagnosed as infected is
presumably harmless, but may be confusing and potentially distort
statistics about risk events.
In Tamarin, soundness is expressed as follows (see \cref{sec:formalisation}
for details):
\begin{lstlisting}
All idR instClose dayClose tsRisk #tRisk.
    PClaimAtRisk(idR, dayClose, instClose)@tRisk
  & Day(tsRisk)@tRisk ==>
(Ex idI place dayContag dayTest [#t1..t6].
    IsAt(idR, place, instClose)@t1
  & IsAt(idI, place, instClose)@t2
  & HAClaimInfected(idI, dayContag, dayTest)@t3
  & EarlierDayEq(dayContag, dayClose)@t4
  & EarlierDay(dayClose, dayTest)@t5
  & Within14Days(dayContag, dayTest)@t6
  & not (idR = idI))
\end{lstlisting}

Whilst the security of the authorisation protocol is implicitly
captured by our soundness property, it is instructive to make it
explicit, as it describes a common point of failure.
In ENS using the \gaen{} framework, the authorisation is particularly sensitive,
as it concerns the publication of
a phone's secret key via the back end.

\begin{definition}[Upload Authorisation for
    \gaen{} protocols]\label{def:upload-auth-dptt}
	If a phone $\id_I$ generated a key $k$ and the back end $B_{cc}$ releases $k$,
        then $\id_I$ was diagnosed as infectious by the health authority.
\end{definition}

In \robert{}, the information disclosed does not originate from the
infected phone, but was picked up by it.

\begin{definition}[Upload Authorisation for
    \robert{}]\label{def:upload-auth-robert}
    If the back end $B_\mathit{cc}$ accepts the upload of ephemerals, presumably
    recorded by a phone $\id_I$,
    then $\id_I$ was diagnosed as infectious by the health authority.
\end{definition}

\subsection{Evaluation Methodology}\label{sec:eval-method}

We seek to systematically characterise the space of possible attacks
from an attacker endowed with a rich set of capabilities that we will
detail below.
As Tamarin is capable of automatically deducing attacks as well as
proofs of their absence, we use it to guide our search by counterexamples, i.e. attacks.
This technique was used in the past to compute strongest threat models
under which a claim holds true~\cite{girolSpectralAnalysisNoise} or
compute protocol security hierarchies~\cite{basinKnowYourEnemy2014}.

Note that both soundness and upload authorisation are
implications of the form $A \implies B$, hence we begin by attempting to
prove $A \implies B$ (`the property holds').
When Tamarin inevitably returns an attack, we (manually) identify an
attack pattern $P_1$, characterising a class of similar attacks using
the compromise events in \cref{tab:adversary} in the Appendix.
We then extend the conclusion to include the possibility of the
attack, i.e. $A \implies B \lor P_1$ (`the original property holds or
the attacker followed $P_1$').
This is a weaker property and thus potentially provable.
We repeat this process until no more attacks are discovered and
Tamarin is able to prove a result of the form
$A \implies B \lor P_1 \lor \cdots \lor P_n$.
\begin{full}
We give a concrete example in \cref{sec:example}.
\end{full}
As we now have a formal proof that there is no attack against
$A \implies B$ unless it matches one of the $P_i$, we obtain an
exhaustive categorisation of all attacks (against this target property
and within the bounds of our formal model).

\begin{nnew}
Compare this with the classical approach, where
a purportedly realistic
threat model is fixed a~priori.
All too often, 
the literature considers only those threat models 
realistic
for which either security can be proven or for which
attacks can be mitigated by technical means, like modifying the protocol.
For
\robert{}, e.g. back end compromise would be excluded from the start, as
there is no way to mitigate this attack.

Similar to a classical threat model, attack patterns give a conditional security
statement: Their negation
$(\neg P_1 \land \ldots \land \neg P_n)$
guarantees the property.
If one thinks that
$(\neg P_1 \land \ldots \land \neg P_n)$
is realistic, then
the proof for
$A \implies B \lor P_1 \lor \cdots \lor P_n$
gives assurance, just like a classical threat model.

This method has three advantages over classical threat models. First, attack
patterns are more precise than threat models typically are: Instead of saying
`the attacker can compromise phones' they can say: `the attacker can compromise one
phone that was in proximity to the victim's phone in the critical time window.'\footnote{%
This is in contrast to prior work
\cite{girolSpectralAnalysisNoise,basinKnowYourEnemy2014},
where attack patterns were limited to adversarial capabilities, which
allowed automating their discovery.
We purposefully explore the space by hand, adding information to the
attack pattern until it is descriptive about the attack.
}
Second, when we consider the necessary condition for the property to hold,
i.e. the above negation $(\neg P_1 \land \ldots \land \neg P_n)$,
we obtain a negative statement about
the attacker (the attacker cannot do X), whereas a threat model gives
a positive statement (the attacker can [only] do Y). It is easier for the
system operator to check if $P_1$ is really impossible than to ensure the threat
model is not missing real-world possibilities. Third, if the attack patterns are
fine-grained enough, their risk can be estimated, whereas the risk `attack
outside the threat model is possible' is harder to quantify.\end{nnew}

\subsection{Adversary Capabilities}

We consider an arbitrary number of phones, with an arbitrary movement and broadcast pattern determined by the adversary.
Key rotation is also under the control of the adversary.
Any number of these phones may go through the testing procedure, begin to upload their keys or finish that process.
Likewise, any combination of phones can download updates from their back-end server and process risk calculations.
We support arbitrary interoperation, as there may be several back ends and health authorities whose regions of coverage may be disjoint or overlapping; however, honest phones interact only with the health authority they were registered with.

The adversary is capable of compromising any of the agents in the system, including the back end infrastructure of the ENS.
They may also compromise a phone, including its key material and records of previously witnessed ephemerals.
Compromising the back end grants the adversary access to the underlying database and the back end's signing key.
Further, whilst the attacker is considered to have full control over messages sent over the Internet, we model the various secure channels between phones, back end and health authority faithfully.
Unless the attacker compromises a device in a particular channel, they are unable to inject or edit messages over the channel, although they may delay or block such messages arbitrarily.
In addition to compromising phones and using them to send ephemerals from particular locations, the adversary may also inject, read and edit Bluetooth messages.
We allow phones and the adversary to be in multiple locations during the same epoch, to reflect the fact that our locations may be very small due to the small intended radius of the Bluetooth protocol.
This allows the adversary to perform relay and replay attacks.
We list all the adversary’s capabilities in \cref{tab:adversary} in the Appendix.

\section{Analysis Results}\label{sec:results}

Based on the evaluation methodology in the last section, we present
a categorisation of all attacks (in our model)
against soundness and upload authorisation
in
\robert{}, \dptt{} and the \cwa{}.

\begin{table}[t!]
    \centering
    \caption{Runtime and memory consumption}
    \label{tab:verification-time}
        \begin{tabular}{@{}lrrr@{}}
\toprule
            &                      \robert{} & \dptt{} & \cwa{} \\ \midrule
Verification time &
  \SI{14}{\hour} \SI{8}{\minute} %
& \SI{1}{\hour} \SI{48}{\minute} %
& \SI{1}{\hour} \SI{3}{\minute} %
                                   \\
Peak memory         &
\SI{43.15}{\giga\byte} & \SI{6.83}{\giga\byte} & \SI{7.67}{\giga\byte}
\\
Proof steps \\
$\quad$Upload auth. &
80 & 75 & 4271 \\
$\quad$Soundness            &
170,137 & 31,561 & 17,262 \\
            \bottomrule
\end{tabular}
\end{table}

\begin{nnew}
We found
27
attack patterns in total\processifversion{full}{, which we present in detail in \cref{sec:full-attacks}}.
\cref{tab:compare} provides a comparative summary of the attacks,
listing the requirements for mounting attacks against upload
authorisation and soundness.%
\begin{full}
        \Cref{tab:attack-surface} provides a more detailed summary.%
\end{full}
The executive overview in \cref{tab:overview} provides an
assessment of the potential impact (in terms of falsely
alerted users) by attack vector.
\begin{conf}
        We present the attacks in full detail in \appendixorfull{sec:full-attacks}.%
\end{conf}
\end{nnew}

\begin{table*}[t!]
    \begin{nnew}
    \small
    \centering
    \begin{tabularx}{\textwidth}{llc@{}cc}
\toprule
                                   & Attack vector                             & \robert{}                                    & \dptt{}           & \cwa{} \\
\midrule
\multirow{2}{*}{Back end}          & \cellcolor{red!25}all-out                 & \textbf{all phones}                          & \multicolumn{2}{c}{all phones in proximity to some other phone} \\
                                   & \cellcolor{red!25}targeted                & any phone                                    & \multicolumn{2}{c}{all contacts of targeted phones} \\
\midrule
\multirow{2}{*}{Health authority}  & \cellcolor{red!10}active                  & any phone                                    & \multicolumn{2}{c}{all contacts of targeted phone} \\
                                   & \cellcolor{red!40}passive (steal QR code) & \textbf{all contacts of arbitrary phone$^*$} & \multicolumn{2}{c}{(phone-and-test-specific authentication codes)} \\ 
\midrule
\multirow{3}{*}{Infected phone(s)} & \cellcolor{red!60}modified phone          & \textbf{contacts of the whole group}         & contacts of phone & max.\ contacts among group mem. \\
                                   & \cellcolor{red!40}+ passive antenna       & \textbf{all phones in reach}                 & \multicolumn{2}{c}{\textit{(see row `modified phone')}} \\
                                   & \cellcolor{red!10}+ active antenna        & \textit{see row `modified phone'}          & \multicolumn{2}{c}{\textbf{all phones in reach}} \\ 
\bottomrule
\multicolumn{5}{l}{(*) Affords attack vector "infected phone"}    
\end{tabularx}
\caption{Executive overview: maximum impact by attack vector and
    protocol. The effect with the strictly highest impact is bold if defined.
    Attack vectors are colour-coded by attack
complexity and detection risk. The more red (greyscale: darker) the attack vector's background, the more realistic it is.}\label{tab:overview}
\end{nnew}
\end{table*}

The verification of the models was performed on an Intel(R) Xeon(R) CPU E5-4650L workstation with four cores for each model.
The verification time, peak memory consumption and proof steps are presented in
\cref{tab:verification-time}. We see that the
majority of the verification time was spent on soundness.

\begin{nnew}
\Cref{tab:compare} and
\xrcite{tab:attack-surface} 
summarise attack patterns by 
target property. For example, the negation of upload authorisation in
\robert{}
(\cref{def:upload-auth-robert})
says that there is a phone $\id_I$ which the back end marks as
infected, but that was not diagnosed as such by the health authority
(similar for
\cref{def:upload-auth-dptt,def:soundness-}).
As we can refer to the target phone $\id_I$ in attack patterns, we can
even include a quantitative assessment. For instance, the attack
pattern \xrref{at:robert-upload-phone} permits \emph{any} phone to
share the QR code received by the health professional, but requires
the \emph{same} phone to upload their EBIDs. For example (simplified):
\begin{lstlisting}
All id t #t1. MarkPositive(id, t)@t1 ==>
    (Ex dt #t3. TestPositive(id, dt)@t3) 
// This was the target property. Now attack pattern 1:
  | (Ex idI ebid qr [..].  
CorruptPhoneReceive(idI, <'infected', [..], qr>)@t1
CorruptPhoneSend(id, <'upload_hello', ebid, [..], qr>)@t2
) | ... // Other attack patterns
\end{lstlisting}
We can test this hypothesis by replacing
$\id_I$,
which represents the infected phone sharing the QR code,
with $\id$, the target phone, in $\tfact{CorruptPhoneReceive}$.
Phone identities in attack patterns are always existentially quantified, 
hence such substitutions strengthen the corresponding lemma.\end{nnew}
If Tamarin can prove this stronger lemma, we report the more specific
attack pattern.
If not, Tamarin provides an attack trace disproving the
modified lemma. As the original lemma was
proven, the attack trace must be the original attack pattern
(otherwise, the original lemma would not be provable)
but the substituted identity must be different from $\id_I$
(otherwise, the modified lemma would be provable).
We manually inspect the attack to see if that phone's identity is
otherwise related to $\id_R$ in particular and if the actions of this
phone can be repeated arbitrarily often.

Besides phone identities, we applied this methodology to QR
codes to find out whether a compromised back end is necessarily
linked to the phone in question.

Previous research has identified many of these attack patterns before. Whilst in some cases, we do discover some new variations that widen their applicability, the differences are not substantial.
Rather, we provably show that these attack patterns characterise all possible violations of the desired security properties and, consequently, rule out further attacks and allow a systematic comparison of the three protocols on a level playing field with a unified terminology.
In these details, we find a number of surprises and previously unremarked differences in attack effectiveness.

\subsection{Executive summary}\label{sec:analysis-summary}

In \cref{tab:overview},
we further condense these patterns to allow for a high-level
comparison of the three designs.

\begin{nnew}
The different effects of a back-end attacker
are a direct and well-known result of the centralised architecture:
In \robert{}, this attacker can simply send a risk notification, while for
\dptt{} and the \cwa{}, Bluetooth communication is required even when the back end is
compromised. If the attack is not targeted, but aims at disrupting the
service, then there is little difference, as we can assume almost any
phone to have been in contact with at least one other phone.

The starkest difference is when a health authority is attacked.
The unspecified upload procedure in \robert{}
seems to have left a weak
spot. A QR code captured in any country is enough to
declare an arbitrary phone infected and thus
trigger the
alarm on a number of phones only limited by how many ephemerals the
attacker can capture, be it by choosing the right phone to declare infected, i.e. one that was near many target victims, or by setting up
an antenna.\end{nnew}

While \dptt{}'s specification achieves strong security properties, the
\cwa{} implements a weaker protocol.\footnote{%
 Note that some real world deployments of \dptt{} also chose a weaker
authorisation scheme.}
The results of our end-to-end analysis on false risk notifications reveals intriguing economics
should the attacker have access to several infected phones.
If a group of pranksters wants to raise
alarms and one of them can obtain a positive test result, then in \dptt{},
they can trigger warnings for everyone who met that person, in the \cwa{}
for whoever in the group has met the most people and in \robert{} for
anyone that met anybody in that group.

\begin{nnew}
The security of authorisation procedure governs how much freedom the
attacker has in choosing the infected phone.
If, in addition, the attacker has the knowledge to set up an antenna,
then the severity of this vector
becomes
a numbers game.

The most effective
attack for a lone individual on \robert{} is to obtain as many ephemerals
as possible and upload them using a compromised phone or a stolen QR code (\xrref{at:robert-relay-malicious}).
Here, the attacker is passive on the Bluetooth channel and
can affect as many phones as the upload limit and the ability
to obtain ephemerals.
Comparable attacks are possible in \dptt{} (\xrref{at:dptt-forge-daily}, \xrref{at:dptt-leak-and-relay-daily}) and the \cwa{} (\xrref{at:cwa-forge-daily}, \xrref{at:cwa-leak-and-relay-daily}), where the attacker uses a disclosed or forged key for broadcasting ephemerals and maliciously uploads it.
However, in \dptt{} and the \cwa{}, the attacks require active transmission on the Bluetooth channel, which risks detection.

\paragraph{Risk Assessment}
While the centralised/decentralised dichotomy has an
impact on back-end compromise, this attack vector requires considerable
expertise and is risky for the attacker.
It is much easier to attack the health authorities, because there are
many of them and they are difficult to secure. In \robert{}, theft of QR codes
can have an enormous impact and requires little expertise. Moreover, health authority's
likely have more than one QR code at their disposal.
By setting up a local instance of the \robert{} back end 
(based on the published source code and its default settings),
we experimentally confirmed 
that a QR code sheet contains QR codes for 10 days,
each with an 8-day validity.

Attacks via modified phones are even simpler to mount and, here again, the
choice between centralised/decentralised comes to effect: If a group
of people decided to disrupt \robert{} in a coordinated effort, they
could easily collect ephemerals and route them to a single phone with
modified software.
The impact can be amplified with an antenna if the attacker has
the expertise.
On our \robert{} instance, we experimentally confirmed that it 
is sufficient to reliably receive ephemerals over a 17-minute period
whereas for \dptt{} and the \cwa{}, the attacker must reliably send for 10 minutes~\cite{baumgartnerMindGAPSecurity2020}.
We consider the risk of a sustained attack with passive and thus
hard-to-detect antennas to be significant and the potential impact high
enough to disrupt the system, at least temporarily.
An active antenna is, nevertheless, a threat if the attacker is risk-tolerant
or intents a one-off attack. \end{nnew}

\begin{table*}[t!]
	\small
	\centering
	\caption{
            High-level comparison of the attack surface. We categorise the
            attack patterns from \protect\appendixorfull{sec:full-attacks}.
			Each bullet point summarises a common attack vector.
	}
	\label{tab:compare}
    \begin{tabularx}{\textwidth}{>{\raggedright\arraybackslash}p{\widthof{Raise alarm on}}>{\raggedright\arraybackslash}X>{\raggedright\arraybackslash}X>{\raggedright\arraybackslash}X>{\raggedright\arraybackslash}X}
		\toprule
		Effect & All & \robert{} & \dptt{} & \cwa{} \\
		\midrule
		Mark $n$~phones infected
		& 
		\begin{tabitemize}
            \item Compromise back end or HA (\xrref{at:robert-upload-qr-list}, \xrref{at:robert-upload-qr-upload}, \xrref{at:robert-upload-secret}, \xrref{at:dptt-upload-other}, \xrref{at:dptt-upload-self}, \xrref{at:cwa-upload-other}, \xrref{at:cwa-upload-self})
		\end{tabitemize}
		& 
		\begin{tabitemize}
            \item Compromise 1 phone at any time (\xrref{at:robert-upload-phone})
            \item Passively compromise HA (obt. QR code) (\xrref{at:robert-upload-qr-list})
		\end{tabitemize}
		&
		\begin{tabitemize}
            \item Compromise $n$ phones prior to testing (\xrref{at:dppt-upload-other-compromised})
		\end{tabitemize}
		& 
		\begin{tabitemize}
            \item Compromise $n$ phones at any time (\xrref{at:cwa-upload-other})
		\end{tabitemize} 
		\\
		Raise alarm on $n$ phones
		&
		\begin{tabitemize}
            \item Relay Attacks (Victim phone $\leftrightarrow$ infected phone) (\xrref{at:robert-relay}, \xrref{at:dptt-relay}, \xrref{at:cwa-relay})
		\end{tabitemize}
		&
		\begin{tabitemize}
            \item Observe $n$ phones and mark $n$ phones as infected (\xrref{at:robert-refl}, \xrref{at:robert-window}, \xrref{at:robert-relay-malicious})
			\item Compromise back end (\xrref{at:robert-false-report}, \xrref{at:robert-backend-qr}, \xrref{at:robert-pubkey}).
            \item Compromise registration phase$^*$ (\xrref{at:robert-pubkey})
		\end{tabitemize}
		&
		\begin{tabitemize}
            \item Replay Attack: Victim $\leftrightarrow$ Positive (\xrref{at:dptt-relay})
			\item Transmit to target and mark $n$ phones as infected (\xrref{at:dptt-mal-upload}, \xrref{at:dptt-window}, \xrref{at:dptt-forge-daily}, \xrref{at:dptt-leak-and-relay-daily}, \xrref{at:dptt-backend-sign}, \xrref{at:dptt-backend-forge-ephs})
		\end{tabitemize}
		& 
	    \begin{tabitemize}
            \item Transmit to target and mark $n$ phones as infected (\xrref{at:cwa-forge-daily}, \xrref{at:cwa-leak-and-relay-daily}, \xrref{at:cwa-backend-forge-ephs}).  
	    \end{tabitemize}
		\\
		\bottomrule
		\multicolumn{5}{l}{(*) The encoding of \cref{def:soundness-} excludes the trivial attack where the phone is fully compromised.}
	\end{tabularx}
\end{table*}

\subsection{Upload Authorisation}

We will now discuss the attacks in more detail, starting with those
against
\cref{def:upload-auth-robert,def:upload-auth-dptt}.

\paragraph{Infrastructure Compromise}
In order for the attacker to list $n$ phones as infected, with all of the systems we studied, it suffices for the attacker to compromise a back end or  health authority (\xrref{at:robert-upload-qr-list}, \xrref{at:robert-upload-qr-upload}, \xrref{at:robert-upload-secret}, \xrref{at:dptt-upload-other}, \xrref{at:dptt-upload-self}, \xrref{at:cwa-upload-other}, \xrref{at:cwa-upload-self}).
In this case, the attacker can send false positive test results or can `sign-off' keys of their choice.
Vulnerabilities of this nature occur in practice, for example, the Swiss contact tracing application 'SwissCOVID', which uses the same design as the \cwa{}, failed to validate signatures made by the health authority~\cite{swisscovidAnnouncementMissingSignature2020}.
Such vulnerabilities can have a much more serious impact on \robert{}, as we will see in the next section.

\paragraph{Device Compromise}
Of course, attacking back-end infrastructure is likely to be much more
difficult than compromising other users’ phones. \robert{}, \dptt{} and the \cwa{}
all achieve very different security properties in this regard. \robert{}
achieves the weakest security property, where the compromise of any
phone that recently has or is about to test positive (or otherwise
obtains a valid QR code) allows the attacker to mark arbitrarily many
devices as infected (\xrref{at:robert-upload-phone}, \xrref{at:robert-upload-qr-list}, \xrref{at:robert-upload-qr-upload}, \xrref{at:robert-upload-secret}).
This is in part because \robert{}, as
specified, has no limitations on the number of ephemerals a user can
upload.\footnote{We did indeed not find any limit on how many ephemerals can
be uploaded in a single batch request in the server source code as
published. In fact, our local tests showed that more than a million EBIDs can be uploaded to our \robert{} server.}
In practice, an operator could review anomalously sized uploads, but the effective upper limit could not be lower than the total number of a user's contacts and is thus likely to be in the hundreds or thousands (e.g. a retail worker in a busy supermarket).
Contrastingly, in \dptt{} and the \cwa{}, a device is limited to uploading a small number of keys, typically 14, since devices only upload their own keys.

\dptt{} achieves a stronger security property than either \robert{} or the \cwa{}, as the attacker must compromise a phone prior to the user getting tested (\xrref{at:dppt-upload-other-compromised}).
This constitutes a substantial barrier to an attacker, since only a small percentage of tested people will actually test positive, meaning the attacker must compromise multiple devices in order to achieve a single malicious upload.
The upload authorisation procedure of \dptt{} requires the user to commit to their device's keys in advance, preventing the attacker from inserting their own keys later.
However, to our knowledge, this authorisation procedure has not been deployed in practice, most likely because it requires a closer integration between the contact tracing system and the health authority's system which may not be practical in all countries.

\begin{nnew}
In the \cwa{}, it is possible to mark people as infected by uploading TEKs of other users (\xrref{at:cwa-upload-other}).
For example, if a TEK has been used by someone near a large group of people which is later uploaded by another positively tested compromised phone.
In practice, however, it can be challenging to extract the TEKs managed by the \gaen{} framework.\end{nnew}

Whilst modelling the \cwa{}, we found that due to underspecification in the upload authorisation procedure, an attacker who learnt the $\guid$ belonging to a phone which had tested positive (e.g. by scanning the QR code sent to the user even after it had been used), could request an arbitrary number of TANs.
We found that this attack had independently been reported to the \cwa{} team~\cite{githubAttackerCanGenerate2020}.
Nevertheless, our analysis proves that enforcing a one-to-one relationship between TANs and registration tokens, as suggested in the initial report, prevents the attack.

\subsection{Raise Alarm on \texorpdfstring{$n$}{n} Phones}

\paragraph{Escaping the Contagious Period}
We discovered a previously unreported attack on soundness in \robert{}:
An attacker who receives a positive test result is able to upload ephemerals outside of their contagious period as determined by a health authority (\xrref{at:robert-window}).
Whilst the attack is relatively minor in isolation, it allows a determined attacker to target victims over a longer period of time than they would otherwise be able to.
For example, whilst the health authority may wish to only begin the tracing period for the previous three days, an attacker would be able to upload ephemerals for the previous two weeks.
\robert{} is vulnerable to this attack, because the authorisation codes used do not identify the start of the contagious period.
Instead, long codes are generated in advance in batches with a validity of 8 days.
We also point out that this long validity period makes authorisation codes more valuable and thus incentivises their compromise.
We have responsibly disclosed this vulnerability to the \robert{} team.

\paragraph{Relay and Replay Attacks}
All three protocols allow an attacker to carry out a relay attack in order to mark honest users as at risk (\xrref{at:robert-relay}, \xrref{at:dptt-relay}, \xrref{at:cwa-relay}).
Whilst the risk of these attacks was generally noted in the early analyses~\cite{aisecPandemicContactTracing2020,dp3tPrivacySecurityAttacks}, the impact of this attack on \robert{} is greater than previously realised.

In this type of attack, the attacker is present at two different locations at the same point in time and forwards messages between them.
In \robert{}, the attacker must observe messages from victims and rebroadcast them near positive individuals.
Conversely, in \dptt{} and the \cwa{}, the direction is reversed: The attacker must broadcast near each victim.
Comparatively, the attacker only needs to broadcast near one infected party in \robert{}.

Real world analysis by \textcite{baumgartnerMindGAPSecurity2020} has shown a substantial asymmetry between transmission and reception of the Bluetooth packets used in \robert{}, \dptt{} and the \cwa{}.
In particular, whilst their tests showed they could effectively receive the Bluetooth packets in real-world conditions (collecting around 20 packets a minute in Frankfurt's central train station), they struggled to transmit reliably at ranges of greater than \SI{10}{\meter} with commodity equipment, e.g. without a dedicated antenna.
Transmission is hampered by the fact that in \dptt{} and the \cwa{} devices only actually wake up to receive messages for a short period of time in a 5-minute window.

This asymmetry makes large-scale relay attacks against \robert{} much more practical than against \dptt{} or the \cwa{}.
In \robert{}, the attacker simply has to acquire a single Bluetooth message from each device it wishes to target which a dedicated attacker can achieve over a large range.\footnote{
	A first estimate targets a maximum attenuation of \SI{92}{\decibel}.
        Assuming line of sight
	and an
	antenna that has \SI{20}{\decibel} gained (e.g. a satellite dish)
	it is plausible to reach a distance of \SI{5}{\kilo\meter}. For comparison, this
	is
	roughly the radius of the city of
	Paris (population 10 M).
	We are not aware of  studies for BLE, but there are
	empirical results on
	802.11b /n/ac~\cite{antonioliPracticalEvaluationPassive2018}, which
	operates on a similar frequency and \SI{20}{\mega\hertz} bandwidth (instead
	of \SI{1}{\mega\hertz}).}
        As mentioned earlier, it is sufficient to 
        reliably receive ephemerals over
        a 17-minute period which corresponds to 102 messages per target
        phone.
Contrastingly, in \dptt{} and the \cwa{}, the attacker must successfully inject multiple messages over a 10-minute period to each target.
Further, the \robert{} attacker is entirely undetectable to its victims, compared to the need to mass-broadcast ephemerals to the \dptt{} and the \cwa{} victims which may attract attention from local authorities and civil fines.
A further distinction is that even when considering a selective attack, a \robert{} attacker may passively gather as many ephemerals as they wish and then decide who should be falsely marked at risk at the last moment.
Contrastingly, a \dptt{} or \cwa{} attacker must pre-commit exactly who they will attack by actively broadcasting ephemerals to them.

\dptt{} is vulnerable to an additional replay attack (\xrref{at:dptt-relay}), previously reported by \textcite{aisecPandemicContactTracing2020} in which the attacker can record an ephemeral and later replay it on the same day.
\dptt{} ephemerals are not bound to specific timeslots, instead they are generated from a single day seed and are broadcast in random order.
Similarly, whilst \cwa{}'s ephemeral identifiers are bound to a specific epoch interval, in practice user devices tolerate packets which may be as much as two hours outside of this interval, ostensibly for reliability when a phone's local time is incorrect.
Contrastingly, \robert{} rejects ephemeral identifiers which are outside a narrow window of a couple of seconds, rendering it invulnerable to this attack.

\paragraph{Infected User Compromise Attacks}
Similar to relay attacks, the adversary may compromise an infected user (via phishing, malware, extortion, bribing or stealing their QR code) in an attempt to have other users alarmed.
That is, unlike in a relay attack, the adversary has privileged access to a device or information belonging to an infected user.

In \robert{}, the attacker must first observe ephemeral broadcasts from the target users, as to inject records of these into the target user’s device (\xrref{at:robert-refl}, \xrref{at:robert-window}, \xrref{at:robert-relay-malicious}).
As in the relay case, the attacker only needs to have access to a single infected user in order to alarm many devices.
With \dptt{} (\xrref{at:dptt-mal-upload}, \xrref{at:dptt-window}, \xrref{at:dptt-forge-daily}, \xrref{at:dptt-leak-and-relay-daily}, \xrref{at:dptt-backend-sign}, \xrref{at:dptt-backend-forge-ephs})
and the \cwa{} (\xrref{at:cwa-forge-daily}, \xrref{at:cwa-leak-and-relay-daily}, \xrref{at:cwa-backend-forge-ephs}),
the attacker must instead transmit to the target users, using ephemerals generated from the user's secret key.

These attacks are especially powerful in \robert{} since the attacker can entirely control how the Bluetooth contact is recorded.
Contrastingly, in \dptt{}, the Bluetooth reception is still performed by an honest device, limiting the attacker to the capabilities of their hardware.
That is, in \robert{}, an attacker could observe a small number of ephemerals in a 15-minute period (e.g. at the very limit of their reception range), but record the contact on the compromised device as a strong, nearby signal for the same duration.
In \dptt{} and the \cwa{}, the attacker would have to be much closer to be able to generate a strong enough signal for the user's device to accept it.

\paragraph{Non-Local Attacks on \robert{}}
For \dptt{} and the \cwa{}, a user can only be falsely alarmed if the attacker is able to either inject Bluetooth messages into their vicinity or compromise a device in the victim's vicinity.
This limited 'blast radius' imposes a substantial economic burden on an attacker wishing to carry out a mass-broadcast attack, as they must be able to inject Bluetooth messages into the proximity of each recipient.
As we noted earlier, this comes with a variety of physical and operational challenges.

Contrastingly, \robert{}'s blast radius is extremely large. An attacker who can compromise the back-end server can alert any and all users using that back end without restriction.
Further, for users of federated \robert{} back ends, the attacker can still cause malicious alerts after observing them once.
Consequently \robert{} is more vulnerable to catastrophic failure than \dptt{} and the \cwa{}.
We do not believe this property has previously been noted.

We also discovered a subtle risk for future implementation
errors: If the back end's secrets become guessable, e.g. due to a bad
key-generation algorithm, the attacker can trigger alarms without
active control over the back end, e.g. without controlling the
channels between back end and phones.
At that point, the attacker can transform any previously received
ephemeral, no matter how old, and convert it into a recent one.
A single infected phone is enough to raise an alarm with any of those.

\paragraph{Federation Attacks on \dptt{} and the \cwa{}}

We find a number of attacks on \dptt{} and the \cwa{} when deployed in a federated setting according to their original specifications.
Whilst we do not report all these attacks here for reasons of space, they arise due to ambiguous requirements regarding the release of the same uploaded key by multiple back ends.
In both systems, the back end must only release a day key once it will no longer be accepted by user devices, for example by waiting until the end of the day the key is reported for.
In the case of the \cwa{} and other systems build on the \gaen{} framework, this necessitates waiting two hours after the end of the day, to account for the clock skew toleration.

In a federated setting, this is compounded since individual back ends may release the same key.
Should any back end release a key 'early' prior to the expiry time of another back end, an attacker can rebroadcast this uploaded key until the later expiry time, triggering false notifications.
A limited form of this attack was reported in~\cite{iovinoEffectivenessTimeTravel2020a} 
with a number of countries; however, we believe the first report \cite{hirshiluccaClockSynchronisationAttack2020} actually dates back to April 2020 by \citeauthor{hirshiluccaClockSynchronisationAttack2020}.
Our analysis shows that ensuring federated servers agree on the expiry date for a particular key provably prevents this class of attacks.
This mitigation has been adopted by the European Federated Gateway Service.

\paragraph{User Compromise}
Another attack possible in any system is when the attacker simply compromises the phone of the victim.
In this circumstance, in \dptt{} and the \cwa{} they can inject a database record corresponding to a past exposure to an infectious person.
In \robert{}, they can simply fake a positive exposure status response from the server.
A similar style of attack was reported by~\cite{iovinoEffectivenessTimeTravel2020a} and dubbed 'time travel’ attacks, as they focused on resetting the phone's clock and simultaneously injecting a Bluetooth message.

\begin{nnew}
\section{Limitations}
\label{sec:limitations}

Our formal analysis is carried out in the Dolev-Yao model where
cryptographic primitives are abstracted, hence attacks against them are not
covered.
Moreover, time and
space are usually not considered in this model. While our spatial model
is reasonably close to reality, our model of time assumes all phones
to have an accurate clock. Hence it cannot capture desynchronisation
attacks (e.g. \cite{perryDevilTimeHow2021}).
In our analysis, we focused on one aspect, the soundness of alarms,
which ensures that every risk notification is justified by a risk
event. However, we do not consider the privacy aspect or completeness
in this work, the latter requiring that every risk event result in a
notification,

We analysed the protocols according to their specification.
We validated how QR codes are distributed in \robert{} on the
published source code, because
this part is underspecified and the impact of the attack
depends crucially on how easy they are to obtain. The other
attacks can be easily validated by specification, hence
a correct implementation would also allow to mount them.
However, we cannot be sure that the implementation does not introduce
new vulnerabilities by diverging from the specification.

All models assume that a particular phone is registered in
only one country and that there is only one back end per
country. 
This is explicit
in the \dptt{} and \cwa{} specifications and implicit in \robert{}'s.

Like the protocol specification, our models abstract both
the health professionals and the infrastructure used to
communicate with the patient into one party, the health
authority.
Our model considers only a single health authority per country, thus
requiring each country's infrastructure to have a single key for
signatures and to share all communication channels.
While this is not strictly required by the specification, in
practice each participating country's infrastructure is set up on the
national level.

Moreover, we assume that a user tests positive only once.
This is a requirement in the \robert{} specification, but the
enforcement mechanism is left abstract. In \dptt{} and the \cwa{},
the specification suggests that all cryptographic material
specific to that device is deleted after testing positive.

We simplify the detailed risk calculation from the model to
a yes/no decision, ignoring the associated encrypted metadata.  
Instead, we allow the adversary to arbitrarily determine the outcome
of risk calculations, which is equivalent to assuming that the
adversary entirely controls the metadata and risk calculation.
Additional model-specific simplifications are detailed in
\cref{sec:modelling-limitations}.\end{nnew}

\section{Conclusion}\label{sec:conclusion}

\begin{nnew}
In this work, we employed formal verification techniques to systematically categorise existing and new attacks on the soundness of risk notifications in ENS.
Although exposure notification protocols seem fairly simple, the details are of utmost importance.
Besides complex temporal and spatial interactions between
phones, cross-country coordination between health authorities,
back ends and authentication providers need to be modelled.
This and the design of fine-grained compromise rules caused our formal models to reach several thousand lines, pushing the practical limit of existing tooling.

Overall, a knowledgeable attacker or
a dedicated group of pranksters could realistically damage a system
with little risk of getting caught.
We want to avoid an oversimplified takeaway message, so we refer to
\cref{sec:analysis-summary} for a half-page executive
summary of the attacks we found.\end{nnew}

Some of the discovered high-level attacks were known to the community (at least
those for \dptt{}), albeit in a simpler form and without a precise
understanding of their consequences and requirements.
We found new attacks that affect possible secondary use cases or
could become relevant due to implementation errors (consider, e.g.
\xrref{at:robert-pubkey}).
We therefore hope that the code which runs on
back-end servers remains open source and is continuously published for
review.

\begin{nnew}
This commendable practice points at an interesting research question:
How can we make sure future versions do not add new attacks? Or that
the code correctly implements the specification.
Extracting protocols from source code is just one possible answer.
Others are monitoring or testing approaches to compare
new versions with formal models such as the ones we presented
here.\end{nnew}

\printbibliography

\appendix

\section{Details on \robert{}}\label{sec:robert-details}

\paragraph{Registration}
The central server is set up with a country code $\CountryCode$, a server key $\SymKeyServer$, a federation key $\FedKeyServer$ and a registration key pair $(\PriKeyServer,\PubKeyServer)$.

After the user downloads the application, the application connects back to the server over TLS and is provisioned with several parameters, such as $\PubKeyServer$, the number of minutes in an epoch and other configuration data.
The application then generates a public and private key pair: $(\PriKeyApp,\PubKeyApp)$. The application sends $\PubKeyApp$ to the server.

The client and server compute their shared secret $\SharedSecretServerApp$ from each other's public key and then derive encryption and authentication keys for the application: $\EncKeyApp,\AuthKeyApp$.

\paragraph{Broadcast}
Immediately after registration and periodically thereafter, the application will connect to the central server in order to obtain a list of encrypted Bluetooth identifiers (EBIDs) and encrypted country code identifiers (ECCs), which the server computes
from its key $\SymKeyServer$ and the federation key $\FedKeyServer$.
\[
    \EBID{A}{\epoch} = \sencBra{\SymKeyServer}{\epoch,\id_A}
    \quad
    \ECC{A}{\epoch} = \sencBra{\FedKeyServer}{\CountryCode,\EBID{A}{\epoch}}
\]
where $\epoch$ is the epoch number.
As $\epoch$ has a bit length of 24 and $\id_A$ of 40 $\EBID{A}{\epoch}$ is produced by
a block cipher with a block size of 64 bits and a key size of 192 bits.
In contrast, the encryption of $\ECC{A}{\epoch}$ uses $\EBID{A}{\epoch}$ as the
initialisation vector
in AES output feedback mode, to link both together.\footnote{The input
    ($\CountryCode$) is actually smaller than one block, so this amounts to
    applying AES to $\EBID{A}{\epoch}$ (with some padding) and xoring the first
8 bits with $\CountryCode$.}

The server encrypts this list using key $\EncKeyApp$ and the application periodically retrieves it.

In each epoch, the phone concatenates $(\EBID{A}{\epoch},\ECC{A}{\epoch})$ with the
lowest 16 bits of the current timestamp $T$. Consequently $T$ resets roughly every 18 hours. This payload is then authenticated with a MAC:
\begin{align*}
    M_{A,\epoch} & = \ECC{A}{\epoch},\EBID{A}{\epoch},T
    \\
    \mathit{MAC}_{A,\epoch} & = \MAC{\AuthKeyApp}{M_{A,\epoch}}
\end{align*}

The resulting pair of messages and MAC are then periodically broadcast.
Any phone in proximity parses the message and checks whether the timestamp is
within a configurable tolerance of the current time (`typically
a few seconds'~\cite{privaticsteaminriaROBERTROBustPrivacypresERving}). Note that $T$ does not prevent inter-day replays, since the counter periodically resets.

\paragraph{Authorisation of Uploads}
When the health authority
diagnoses the owner of $P$ as
contagious, a contagious period is determined.
The health authority then
proposes the owner to upload all ephemerals that it collected from other
phones during this time period to the server,
along with the time that it received them.
The specification explicitly leaves the authorisation procedure open, but
requires that `only authorised and positively tested users are allowed
to upload'.
We thus investigated the source code of the back-end server, submission code
server and the REST API specification.%
\begin{nnew}ROBERT has \emph{long codes} which are encoded as QR codes and distributed
via the SIDEP platform, a centralised system for registration of test results and are valid for 8 days.
Patients that are not confirmed via a SIDEP-connected laboratory
or diagnosed as a probable case by a health professional obtain
\emph{short codes} which are valid for only 60 minutes.
The SIDEP platform allows producing long codes in batch, hence the
health professional can obtain long codes with starting dates up to 10 days in the
future (and aforementioned 8 days of validity).\end{nnew}

The phone obtains such a code from a health authority as a
QR code and attaches it to the ephemerals that it wants to upload.
The server
(1) validates the token
and
(2) parses the ephemerals. For each transmitted ephemeral, the server
(3) checks that the reception time and the
16-bit timestamp in the ephemeral roughly match,
(4) decrypts $\ECC{A}{\epoch}$ and compares it with its own country code,
forwarding the message to the right server, if necessary.
Then, (5) $\EBID{A}{\epoch}$ is decrypted and parsed to $(\epoch,\id_A)$,
(6) $\id_A$ checked to belong to a registered phone,
and
(7) $\epoch$ compared with the claimed reception time.
With $\id_A$, (8) the emitting phone's
key is retrieved and (9) used to validate the MAC.
Finally, $\epoch$ is added to the database $\LEE_{\id_A}$ as an epoch where $\id_A$ was
exposed ($\LEE$ stands for list of exposed epochs).

As the identity assigned to each phone is in principle opaque, the \robert{} specification suggests that additional actions, such as the deployment of a mixnet, could be taken to prevent the server from learning the uploader's identity through metadata such as IP addresses.
In practice, no such system has been deployed.

\paragraph{Risk Calculation}
The risk of exposure is computed by the back end upon request by the
phone.
This request contains
the current epoch $\epoch$, time
$t_\mathvalue{req}$ and
$\EBID{A}{i}$, $\ECC{A}{i}$.
The EBID and ECC values are used to authenticate the origin of this request,
which is performed exactly with the same steps (1) to (8) as above.

\begin{full}
Some additional checks ensure that the user does not query too
frequently. We ignore them, as we do not aim to establish
availability.
\end{full}
If the authenticity of the origin is established, the server returns `1' in case
the risk score
exceeds some
threshold, and `0' otherwise.
The risk score is
derived from this phone's ($\id$'s) list of exposed epochs $\LEE_\id$
and metadata like
reported signal strength and duration,

\paragraph{Federation}
Interoperability between back ends is achieved using the
shared federation key $\FedKeyServer$.
Each  back end–-there is one per country code $\CountryCode$
---forwards the recorded EBID and its metadata on a secure channel to the back end identified by
the encrypted country code $\ECC{A}{\epoch}$.
The receiving back end then processes the record as normal.

\section{Details on \gaen{}}\label{sec:gaen-details}
\paragraph{Broadcast}

The \gaen{} framework divides time into consecutively numbered 10 minute epochs since the start of Unix time.
When the framework is first activated on a device, a fresh Temporary Exposure Key (TEK) is generated and tagged with the current epoch index: $\TEK{d}$.
After the end of each 24-hour period, a fresh TEK is generated for use in that period and any TEKs older than 14 days are deleted.

In a particular 24-hour period with $\TEK{d}$, two further keys are derived using HKDF:
\begin{align*}
    \RPIK{d} & = \HKDF{\TEK{d}, ``\mathtt{ENRPIK}"} \\
    \AEMK{d} & = \HKDF{\TEK{d}, ``\mathtt{ENAEMK}"}
\intertext{%
The RPIK and AEMK are then used to derive a Rolling Proximity Identifier (RPI) for each epoch $j$ and some encrypted associated metadata:
}
    \RPI{d}{j} & = AES_{128}(\RPIK{d},j) \\
    \AEM{d}{j} & = AES_{128}(\AEMK{d},j,``\mathtt{Metadata}")
\end{align*}

The metadata contains the version of the framework used by the transmitter and the estimated transmission power.
The device then periodically broadcasts these two values and records any ephemerals it receives from nearby devices, alongside the reception time and relative transmission strength.

\paragraph{Risk Calculation}

The framework handles risk calculation on behalf of the contact tracing application.
Indeed, the framework (in conjunction with the operating system) actively protects the stored keys and recorded ephemerals.
However, the contact tracing application can pass a list of newly uploaded keys to the framework which will then generate the resulting identifiers and check whether it has observed any.
This matching process has a two-hour tolerance to allow for clock skew on the sending or receiving device.

As a further security measure, newly provided keys must be signed by the operator of the national back end.
The signature public key is certified by Google/Apple.
\section{Details on the \cwa{}}\label{sec:cwa-details}

\begin{figure}[!ht]
	\centering
	\input{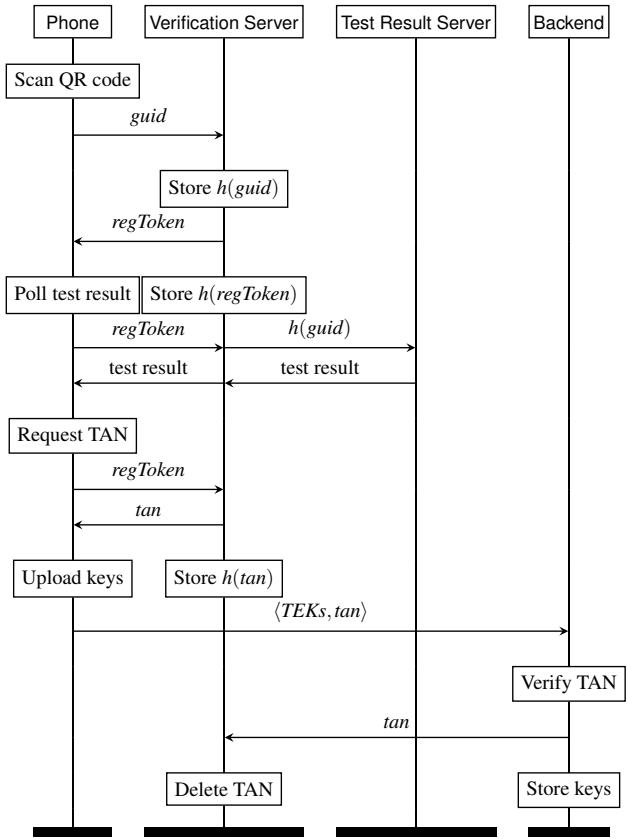}
	\caption{Authorisation procedure using a QR code for the \cwa{}, where $\guid$, $\cwatan$ and $\rg$ are freshly generated.}
    \label{msc:cwa-auth-qr}
\end{figure}

\paragraph{QR Code Based Upload Authorisation}
This protocol is depicted in \cref{msc:cwa-auth-qr}.
When a test is conducted by a health professional, the user scans a QR code with their phone which contains a randomly chosen $\guid$.
The phone hashes this $\guid$ and sends it to a verification server (VS) to obtain a \emph{registration token} $\rg$.
The verification server stores the hashed $\guid$ and hashed registration token $\rg$.
The phone then polls in a regular interval for updates on its owner's test results by sending its registration token to the VS.

The hashed $\guid$ is also attached to the test sample analysed by the laboratory.
The VS can thus request updates on a test result by sending the hashed $\guid$ to a test result server (TRS).
The next time the phone polls, the verification server will forward the test result from the TRS.

If the test result is positive, the phone can request a TAN from the VS, which stores a hashed version of the TAN.
The phone can then start the upload process by attaching the TAN to its TEKs.
The back end will check whether the TAN is valid by contacting the VS in which case the hashed TAN is deleted from the VS.

\paragraph{Telephone Based Upload Authorisation}
As some testing facilities are not equipped to provide QR codes and some patients may fail to scan the QR code at the time, a backup authentication protocol is also offered.
In this protocol, the user calls a medical hotline, is interviewed to ensure they recently had a positive test and receives a teleTAN directly over the phone which they enter into their application.
The phone then contacts the verification server, provides the teleTAN and receives a registration code.
The rest of the protocol proceeds as in the QR code variant.

\paragraph{European Federation Gateway Service}
In the \gaen{} framework, each country must provide its own back end and national infrastructure.
This raises a question of how to handle travellers between countries, who are typically a high-risk category for COVID-19 exposure.
The European Federation Gateway Service (EFGS) was developed by T-Systems to provide a solution across the EU.
As detailed in~\cite{ehealthnetworkEuropeanProximityTracing}, the EFGS architecture uses a central database server which every country's back end connects to via mutually authenticated TLS.

Users who are uploading their keys are expected to mark which countries they have been present during their contagious period.
Their back end includes this information when it uploads their keys to the EFGS database.
Periodically, the other back ends fetch all new keys from this database and build a list of contagious keys for each country in the federation.
Users then periodically download all the keys corresponding to countries they have recently visited.

\begin{full}
    \section{Full Attack Description}\label{sec:full-attacks}
    \begin{table*}[!ht]
    \centering
    \caption{%
        The attack surface exhibited by \robert{}, \dptt{} and the \cwa{} showing
        what the attacker needs to do to achieve the desired
        effect.
        Listed are the requirements for attacks in
        \Cref{sec:full-attacks}.
        Each bullet point corresponds to one or more distinct attacks.
        Hence, they need to be read as a disjunction.
        We omit attacks dominated by other attacks
        that have the same effect but stronger requirements. 
}
    \label{tab:attack-surface}
    \small
    \begin{tabularx}{\textwidth}{>{\raggedright\arraybackslash}p{\widthof{Violation of Def. X}}>{\raggedright\arraybackslash}X>{\raggedright\arraybackslash}X>{\raggedright\arraybackslash}X}
        \toprule
        Effect & \robert{} & \dptt{} & \cwa{}
        \\\midrule
        Violation of
    Def.~\ref{def:upload-auth-dptt} + Def.~\ref{def:upload-auth-robert}:
Mark $n$~phones infected
               & 
\begin{tabitemize}
    \item Obtain 1\maxuploadf QR code\anycc and upload with \emph{these} phones  (\ref{at:robert-upload-phone}, \ref{at:robert-upload-qr-list}, \ref{at:robert-upload-qr-upload})
\item Compromise $1$ back end\anycc (\ref{at:robert-upload-secret})

\end{tabitemize}
               &
\begin{tabitemize}[nosep]
    \item Compromise \emph{these} $n$ phones and $1$ health authority\anycc (\ref{at:dptt-upload-other}, \ref{at:dptt-upload-self})
\item Compromise \emph{these} $n$ phones and let $n$ compromised infected phones upload their keys (\ref{at:dppt-upload-other-compromised})

\end{tabitemize}
               &
\begin{tabitemize}[nosep]
    \item Compromise $n$ phones\samecc or their back end (\ref{at:cwa-upload-other})
\item Compromise $1$ verification server\samecc and either $1$ phone\samecc or $1$ back end\samecc (\ref{at:cwa-upload-self})

\end{tabitemize}

\\\\\\
Violation of Def.~\ref{def:soundness-}:
Raise alarm on $n$ phones
               &
\begin{tabitemize}[nosep]
    \item Compromise $1$ (or many) infected phones and, during the last 14 days\outsidef, place in proximity to all $n$ phones (\ref{at:robert-refl}, \ref{at:robert-window})
\item Obtain\obtainf $n$ ephemerals\dayf and relay to 1\maxuploadf infected phone\anycc (\ref{at:robert-relay})
\item Obtain\obtainf $n$ ephemerals\dayf and let 1\maxuploadf compromised infected phone upload\anycc  (\ref{at:robert-relay-malicious})
\item Compromise $1$ back end\samecc (active attack, \ref{at:robert-false-report})
\item Steal $1$ back end key\samecc + 1 federation key and compromise 1 infected phone\samecc (\ref{at:robert-backend-qr})
\item Learn/inject $n$ phones' public keys (\ref{at:robert-pubkey})

\end{tabitemize}
               &
\begin{tabitemize}[nosep]
    \item Compromise $1$ (or many) infected phones and, during the last $14$ days\outsidef, place in proximity to all $n$ phones (\ref{at:dptt-mal-upload})
\item Compromise $1$ infected phone and, within contagious period, reach all $n$ phones via Bluetooth (\ref{at:dptt-window})
\item Relay $1$ infected ephemerals\dayf to $n$ phone(s)\anycc (\ref{at:dptt-relay})
\item Compromise $1$ phone and $1$ health authority\anycc and, before malicious upload, reach all $n$ phones via Bluetooth (\ref{at:dptt-forge-daily}, \ref{at:dptt-leak-and-relay-daily})
\item Compromise $1$ back end\samecc and, before upload, reach each phone via Bluetooth (\ref{at:dptt-backend-sign}, \ref{at:dptt-backend-forge-ephs})

\end{tabitemize}
               &
\begin{tabitemize}[nosep]
    \item Relay $1$ infected ephemerals\dayf to $n$ phone(s)\anycc (\ref{at:cwa-relay})
\item Compromise $1$ phone and either $1$ back end\samecc or $1$ verification server\samecc and, before malicious upload, reach all $n$ phones via Bluetooth (\ref{at:cwa-forge-daily}, \ref{at:cwa-leak-and-relay-daily})
\item Compromise $1$ back end\samecc and, before upload, reach each phone via Bluetooth (\ref{at:cwa-backend-forge-ephs})

\end{tabitemize}
\\
        \bottomrule
    \end{tabularx}

\footnotesize
\begin{tabularx}{\textwidth}{l@{\hskip 4pt}c@{\hskip 4pt}X}
    \anycc $^{/}$ \samecc  & = & For any/same country's back end.
 \\
    \dayf & =& Ephemerals for same day (\robert{}: same 18-hour window), but emitted at possibly different place.
    \\
    \obtainf &=& Ephemerals can be obtained by
    \begin{enumerate*}[label=(\alph*)]
        \item recording them via Bluetooth,
        \item compromising a phone that received them via Bluetooth,
        \item capturing the so-called pre-Hello message sent from the,
        back end to the target phone, or
        \item capturing a status request to the back end.
    \end{enumerate*}
    \\
    \outsidef & =& Possibly outside the contagious window.
\end{tabularx}
\end{table*}

    \subsection{Security of Upload Authorisation} 

    \paragraph{\robert{}}

When a phone breaks upload authorisation, all users whose ephemerals
were in the upload message will receive a risk notification from the back end
even though they have not necessarily been in contact with an infected person.

There are four different attack patterns against the security of upload
authorisation in \robert{}. Each of these attacks is characterised by
a compromised phone uploading an ephemeral with an illegally obtained
QR code thereby authorising the upload.
The uploaded ephemeral is known to the attacker.

The QR code is obtained either
\begin{enumerate}[label={(A\arabic*)},ref={A\arabic*}]
    \item \label{at:robert-upload-phone}
        from any compromised phone, possibly registered with a different
        back end, which is tested positive and shares its QR code
        with the attacker;
    \item \label{at:robert-upload-qr-list}
        from a (possibly different) back end that leaks the QR code
        from its internal list of valid QR codes;
    \item \label{at:robert-upload-qr-upload}
        from a (possibly different) back end that received a (possibly honest)
        upload message and leaks the QR code therein; or
    \item \label{at:robert-upload-secret}
        by obtaining the secret key of a (possibly different)
        back end and creating new QR codes (the back end is responsible
        for signing QR codes).
\end{enumerate}

The first case requires the compromise of \emph{infected} phones, which is
hard to do at scale. However, it suffices to compromise infected phones of any country.

The other three cases show that the compromise of a country's back end is
sufficient to attack the registration of users in any other country.
This can be done through the leakage of QR codes (e.g. a sheet with
QR codes is stolen from a practitioner's desk) or by attacking the
back end itself.

In the first three cases, the number of users receiving a false risk notification
is limited by the number of obtained QR codes times the maximal number of allowed
ephemerals in an upload or the number of obtained ephemerals.
In the last case, it is only limited by the number of obtained ephemerals.

    \paragraph{\dptt{}}
\label{p:dp3t-upload}

When a phone breaks upload authorisation, all users who received an ephemeral from the compromised phone will compute a risk score that is positive even though they have not necessarily been in contact with an infected person.

There are three attack patterns against the security of upload authorisation in \dptt{}.
The first two attacks are characterised by a phone disclosing their daily key and a health authority leaking their signing key.
With both keys, the adversary can construct valid ephemerals.
These can be uploaded either
\begin{enumerate}[label={(B\arabic*)},ref={B\arabic*}]
    \item \label{at:dptt-upload-other}
        by another compromised phone (which is not necessarily tested positive), or
    \item \label{at:dptt-upload-self}
        by the original phone where the compromised health authority sends an illegitimate positive test result to the phone to authorise the upload.
\end{enumerate}
In the second case, it is also required for the original phone to leak their nonces stored in their test databases or allow the adversary to write a value of their choice.
Instead of the health authority, the attacker can compromise a phone that tested positive.

The third attack requires the compromise of an infected phone.
\begin{enumerate}[label={(B\arabic*)},ref={B\arabic*},resume]
    \item \label{at:dppt-upload-other-compromised}
        A phone that tested positive can upload another phone's key if both are compromised.
\end{enumerate}
    \paragraph{\cwa{}}
\label{p:cwa-upload}

When a phone breaks upload authorisation, all users who received an ephemeral from the compromised phone will compute a risk score that is positive even though they have not necessarily been in contact with an infected person.

There are two different attack patterns against the security of upload authorisation in the \cwa{} which are characterised by disclosing a valid TAN.
This TAN can either be leaked by a compromised phone or by a compromised back end leaking the TAN used for a (possibly honest) upload of a key.

The upload of the TEK can then happen in two ways.
\begin{enumerate}[label={(C\arabic*}),ref={C\arabic*}]
    \item \label{at:cwa-upload-other}
        A TEK is disclosed either by a phone or by a compromised back end leaking the TEK used for a (possibly honest) upload.
        This TEK is then uploaded by a compromised phone that has not been tested positive at this time.
    \item \label{at:cwa-upload-self}
        The verification server sends an illegitimate positive test result to a phone to authorise the upload.
\end{enumerate}

In the second case, no phone has to be compromised, in contrast to \dptt{}, where the phone's key has to be disclosed since it is included in the test result notification of the health authority.

    \subsection{Soundness} 

There are various attacks against soundness that do not rely on
unauthorised uploads. We categorise them into \emph{phone compromise}
attacks, \emph{relay} attacks and \emph{back end compromise} attacks.

    \paragraph{\robert{}}

There are seven ways of attacking soundness in \robert{}. The
following two attacks rely on infected phones being compromised:
\begin{enumerate}[label={(X\arabic*)},ref={X\arabic*}]
    \item \label{at:robert-refl}
        Reflection attack (not displayed in \cref{tab:attack-surface}).
        According to specification, a user
        may receive their own broadcast
        messages and thus be registered as being at risk---in addition
        to being positively diagnosed earlier.
        This attack may be relevant in case the government is using the
        ENS for statistics about how many users a sick person is
        potentially infecting.
        This attack can be easily thwarted by server side filtering.

    \item \label{at:robert-window}
        Infected phones obtain a contagious window from the health authority, but
        that window is not authenticated---the QR code is just
        a one-time token to upload ephemerals. Hence an infected phone
        can trigger risk notifications for phones it has been in
        contact with in the last 14 days, even if outside the
        contagious period.
\end{enumerate}

Since in \robert{} a single infected phone can mark many other phones
as at risk, there are many relay attacks.
\begin{enumerate}[label={(X\arabic*)},ref={X\arabic*},resume]
    \item \label{at:robert-relay}
        An attacker can obtain ephemerals at one place and replay
        them at another place at the same time.
        If these are picked up by a phone that is later diagnosed as infected,
        the phones sending them receive a false risk notification.
        Note that one infected phone is sufficient.
        Hence, the impact of the attack scales linearly with the number of ephemerals the attacker
        can pick up and forward in time (there is one ephemeral per phone per time unit).%
        In the context of \dptt{},
        Vaudenay called this attack `Relay Attack'~\cite{vaudenayAnalysisDP3T2020}.
        This attack is arguably more severe in \robert{} than in \dptt{} under
        circumstances where emitting Bluetooth ephemerals is more costly
        than receiving them.
        This can be the case if a single attacker aims for a large-scale
        disruption: In \robert{}, the attacker can be passive, as
        just one infected recipient is enough to attack many different
        phones. In \dptt{}, they must be active and broadcast many
        ephemerals, possibly revealing their location.

    \item \label{at:robert-relay-malicious}
        If the infected phone in the previous attack is compromised
        (or can at least upload ephemerals of the attacker's choice),
        then no Bluetooth transmission is necessary.
\end{enumerate}

For both of the above attack patterns, the ephemerals can be obtained
in four ways:
\begin{enumerate*}[label=(\alph*)]
    \item by recording them via Bluetooth,
    \item by compromising a phone that received them via Bluetooth,
    \item by capturing the pre-hello message from the back end to the target phone, or
    \item by capturing a status request to the back end.
\end{enumerate*}

There are two attacks based on back-end compromise.
\begin{enumerate}[label={(X\arabic*)},ref={X\arabic*},resume]
    \item \label{at:robert-false-report}
          A compromised back end can send a false risk notification to any phone
          of the same country.
    \item \label{at:robert-backend-qr}
        When the adversary obtains the federation key and the secret
        key of the victim's back end,
        it can reconstruct ephemerals without having to observe them.
        This attack on the back end is passive and can be executed
        if the back end's key is guessable, e.g. when randomness
        generation is flawed. Furthermore, it cannot be
        detected by the server.
        The ephemerals are then uploaded to the unsuspecting back end
        by a compromised phone.
        To obtain the QR code, any attack on upload authorisation
        (\ref{at:robert-upload-phone}--\ref{at:robert-upload-secret})
        suffices.
        Nevertheless,
        upload authorisation is not necessarily violated, as the phone
        uploading the ephemeral may still be tested positive later.
\end{enumerate}

The last attack does not fall into any of the aforementioned categories.
\begin{enumerate}[label={(X\arabic*)},ref={X\arabic*},resume]
    \item \label{at:robert-pubkey}
           \begin{nnew}If the adversary can inject or guess a target phone's public key,
           or
           learn the pre-HELLO message generated from the back end and
           transmitted to the target phone,
           they can construct new ephemerals and use them to provoke
           an alarm on the target.
           This does not require any Bluetooth communication, any
           number of corrupted helper phones can upload the ephemerals
           without having been in contact and the attacker needs not
           to commit on any specific helper phone becoming infected
           later---one is sufficient. 

           The phone's public key is exchanged during registration,
           the pre-HELLO messages are computed by the back end and
           transmitted during registration and then regularly updated.
           Hence, a vulnerability in the registration process can lead to false risk
           notifications on all registered phones without back-end
           compromise or BLE communication.
           A third, less realistic variant of the attack 
           obtains the secret keys and the federation key of the phone's back end.\end{nnew}

\end{enumerate}

    \paragraph{\dptt{}}

There are seven ways of attacking soundness in \dptt{}.
The following two attacks rely on infected phones being compromised.
\begin{enumerate}[label={(Y\arabic*)},ref={Y\arabic*}]
    \item \label{at:dptt-mal-upload}
        An ephemeral has been exchanged by two phones in proximity in the last 14 days, but not necessarily in the contagious period of the infected phone.
        The ephemeral is maliciously uploaded by the infected phone as described in \cref{p:dp3t-upload}.
        Since the infected phone has been tested positive, upload authorisation is not violated.

    \item \label{at:dptt-window}
        A daily key of an infected phone is disclosed and used by the adversary to construct an ephemeral, which is distributed over Bluetooth.
        The broadcast happened in the last 14 days, but not necessarily in the contagious period of the infected phone.
        Each recipient of the ephemeral receives a false risk notification.
        Since the infected phone has been tested positive, upload authorisation is not violated.
\end{enumerate}

In \dptt{} receiving a single ephemeral of an infected phone is sufficient to receive a risk notification resulting in Vaudenay's relay attack~\cite{vaudenayAnalysisDP3T2020}.
\begin{enumerate}[label={(Y\arabic*)},ref={Y\arabic*},resume]
    \item \label{at:dptt-relay}
        The adversary distributes an ephemeral over Bluetooth which was previously broadcast by an infected phone.
        The adversary learnt the ephemeral either through direct reception or compromise of a phone which received the ephemeral.
        The replay happens at the same day, but not necessarily in the same epoch.
        Since the infected phone has been tested positive, upload authorisation is not violated.
\end{enumerate}

In the following four cases, at least one health authority or back end must be compromised.
\begin{enumerate}[label={(Y\arabic*)},ref={Y\arabic*},resume]
    \item \label{at:dptt-forge-daily}
        The adversary forges daily keys to construct ephemerals, which are broadcast over Bluetooth.
        These forged ephemerals are then maliciously uploaded as described in \cref{p:dp3t-upload}, only that daily keys have to be leaked.
        Since upload authorisation is only specified with respect to honestly generated keys, it always holds in this case.

    \item \label{at:dptt-leak-and-relay-daily}
        The daily key of a phone is disclosed and uploaded maliciously as described in \cref{p:dp3t-upload}.
        The ephemeral can be replayed multiple times at different places on the same day but with possibly different epochs.
        Each recipient receives a false risk notification.
        Since the phone whose key has been disclosed does not have to be tested positive, upload authorisation may be violated.

    \item \label{at:dptt-backend-sign}
        An ephemeral is exchanged by two phones in proximity and the daily key of the sender is disclosed.
        The signing key of the back end is leaked and used to send a false risk notification to the receiver of the ephemeral.
        Since no upload took place, upload authorisation is not violated.

    \item \label{at:dptt-backend-forge-ephs}
        The adversary distributes (possible forged) ephemerals over Bluetooth and compromises a back end to send false risk notification to all recipients of the broadcast ephemeral.
        Only the back end has to be compromised.
        Since no upload took place, upload authorisation is not violated.
\end{enumerate}
    \paragraph{\cwa{}}

There are four ways of attacking soundness in the \cwa{}.
Receiving a single ephemeral of an infected phone is sufficient to receive a risk notification, resulting in Vaudenay's relay attack~\cite{vaudenayAnalysisDP3T2020}.
\begin{enumerate}[label={(Z\arabic*)},ref={Z\arabic*}]
    \item \label{at:cwa-relay}
        The adversary distributes an ephemeral over Bluetooth which was previously broadcast by an infected phone.
        The adversary learnt the ephemeral either through direct reception or compromising a phone which received the ephemeral.
        The replay happens on the same day but not necessarily in the same epoch.
\end{enumerate}

In the following three cases, at least one health authority or the back end must be compromised.
\begin{enumerate}[label={(Z\arabic*)},ref={Z\arabic*},resume]
    \item \label{at:cwa-forge-daily}
        The adversary forges daily keys to construct ephemerals which are broadcast over Bluetooth.
        These forged ephemerals are then maliciously uploaded as described in \cref{p:cwa-upload}, only that daily keys have to be leaked.
        Since upload authorisation is only specified with respect to honestly generated keys, it always holds in this case.

    \item \label{at:cwa-leak-and-relay-daily}
        The daily key of a phone is disclosed and uploaded maliciously as described in \cref{p:cwa-upload}.
        The ephemeral can be replayed multiple times at different places on the same day but with possibly different epochs.
        Each recipient receives a false risk notification.
        Since the phone whose key has been disclosed does not have to be tested positive, upload authorisation may be violated.

    \item \label{at:cwa-backend-forge-ephs}
        The adversary distributes (possibly forged) ephemerals via Bluetooth and compromises a back end to send false risk notification to all recipients of the broadcast ephemeral.
        Only the back end has to be compromised.
\end{enumerate}

\end{full}

\begin{table*}[t]
    \begin{nnew}
    \caption{Adversarial Capabilities}
    \label{tab:adversary}
    \small
    \centering
    \begin{tabularx}{\textwidth}{lXp{\widthof{$\quad\mathsf{CorruptVS(SendTo|ReceiveFrom)TRSnB}, m$}}}
		\toprule
		Infrastructure         & Capabilities                                                      & Events \\
        \midrule
        Internet               & Eavesdrop and inject messages                                     & \textit{built-in} \\
        \midrule
        Bluetooth              & Read and write message $m$ at time $t$ and location $p$           & $\tfact{BLErd}(t, p, m)$, $\tfact{BLEwr}(t, p, m)$ \\
        \midrule
        \multirow{6}{*}{Phone} &                                                                   & $\tfact{Corrupt}(\langle 'P', id \rangle, \star)$ with $\star=$ \\
                               & Reveal long-term (\robert{}) or day key $k$ (\dptt{}, \cwa{})     & $\quad\mathsf{CorruptPhoneKey}, k$ \\
                               & Reveal witnessed ephemeral $\eph$                                 & $\quad\mathsf{CorruptPhoneReceived}, \eph$ \\
                               & Send and receive message $m$ to and from back end or HA           & $\quad\mathsf{CorruptPhone(Send|Receive)}, m$ \\
                               & Read and write testing state $s$ (\dptt{})                        & $\quad\mathsf{CorruptPhoneTestDB(Read|Write)}, s$ \\
        \midrule
        \multirow{10}{*}{Back end} &                                                          & $\tfact{Corrupt}(\langle 'B',\cc \rangle, \star)$, with $\star=$ \\
                                   & Reveal long-term key $k$                                 & $\quad\mathsf{CorruptBState}, k$ \\
                                   & Receive message $m$ from phone (\robert{}, \dptt{})      & $\quad\mathsf{CorruptBReceive}, m$ \\
                                   & Send message $m$ to VS (\cwa{}) or phone (\robert{})     & $\quad\mathsf{CorruptBSend}, m$ \\
                                   & Receive message $m$ from VS (\cwa{})                     & $\quad\mathsf{CorruptBReceiveFromVS}, m$ \\
                                   & Receive message $m$ from phone (\cwa{})                  & $\quad\mathsf{CorruptBReceiveFromPhone}, m$ \\
                                   & Reveal QR codes $\qr$ (\robert{})                        & $\quad\mathsf{CorruptQRList}, \qr)$ \\
                                   & Reveal phone registration keys and identifer (\robert{}) & $\quad\mathsf{CorruptBIDTable}, (k_1,k_2,\id)$ \\
                                   & Reveal federation key $k$ (\robert{})                    & $\quad\mathsf{CorruptBFederationKey'}, k$ \\
        \midrule
        \multirow{3}{*}{Verification server (VS)} &                                                                  & $\tfact{Corrupt}(\langle 'VS',\cc \rangle, \star)$, with $\star=$ \\
                                                  & Send and receive message $m$ to and from back end (\cwa{})       & $\quad\mathsf{CorruptVS(SendTo|ReceiveFrom)TRSnB}, m$ \\
                                                  & Send and receive message $m$ to and from phone (\cwa{})          & $\quad\mathsf{CorruptVS(SendTo|ReceiveFrom)Phone}, m$ \\
        \midrule
        \multirow{3}{*}{Health authority (HA)} &                                     & $\tfact{Corrupt}(\langle 'HA', \cc \rangle, \star)$, with $\star=$ \\
                                               & Reveal long-term key $k$ (\dptt{})  & $\quad\mathsf{CorruptHAState}, k)$ \\
                                               & Send message $m$ to phone (\dptt{}) & $\quad\mathsf{CorruptHASend}, m)$ \\
        \bottomrule
	\end{tabularx}
    \end{nnew}
\end{table*}
\section{Modelling Challenges}\label{sec:modelling-challenges}

While the high-level description of the protocols is simple, we aimed
for a detailed analysis that includes authentication procedures and
federation, leading to a large model size (several thousand lines) and
12 person months of model development. Behind the models
of the
TPM~\cite{10.1145/3320269.3372197},
5G~\cite{basin2018formal,peltonen2021comprehensive},
DNP3~\cite{cremers2019secure}
and
ISOIEC~\cite{whitefield2019symbolic},
the \cwa{} model (1053 lines) appears to be the fifth-largest protocol
model
in Tamarin's model repository,\footnote{We count variants of the same
protocol (e.g. there are two models for
5G~\cite{basin2018formal,peltonen2021comprehensive}) only once.
Furthermore, there are models not contained in Tamarin's repository,
e.g. TLS 1.3~\cite{cremers2016automated}.}
with the \robert{} and \dptt{} models being
only slightly smaller (935 and 832 lines).

In contrast to traditional analyses, we developed a granular
compromise analysis, which meant that certain simplifications were
unavailable (e.g. combining the store of the health authority and the back end into
one to avoid modelling their communication) and that Tamarin's
analysis had to regard more edge cases. This, in turn, required us to
write custom
heuristics that prioritise resolving goals that help conclude from the
occurrence of certain messages to which parties need to be compromised
to produce this message.

We already described our evaluation methodology in
\cref{sec:eval-method}, which we believe to be a novel approach to
formal analysis, exploring the space of compromise scenarios rather
than setting in a~priori, which in this case was simply not possible,
because they were not yet known. We exemplify the methodology
below, but first discuss some modelling aspects.

\subsection{Modelling Oracles}\label{sec:oracles}

The oracles we developed are relatively simple in operation, but
required manual inspection of the proof tree to be developed.
The two main ideas are
\begin{enumerate*}[label=(\alph*)]
  \item prioritising constraints involving the adversary deducing a private key, as these typically rapidly lead to contradictions; and
  \item deprioritise constraints involving the ordering of particular events, as those lead to additional cases and grow quadratically in the number of distinct time points, without progressing the analysis in terms of message deduction.
\end{enumerate*}

\begin{nnew}
\subsection{Modelling Limitations}
\label{sec:modelling-limitations}

\paragraph{\robert{}}

The \robert{} specification requires that the back end notifies a user that they are at risk only once.
Our model reflects this requirement, although we
are unsure if a practical system deployed at scale could successfully
maintain this invariant.

\paragraph{\dptt{}}

In \dptt{}, we assume that uncompromised phones only generate a single TEK on a given day.
In certain rare situations this may not be the case, but our analysis also covers compromised phones which may use multiple TEKs simultaneously.
When phones upload their key material, we model this as multiple independent uploads rather than a single upload containing a list of keys.
This simplifies the analysis but does not change the behaviour.

\paragraph{\cwa{}}

We make similar assumptions as for the \dptt{} model.
In the upload authorisation procedure, we model the verification server and the test result server as one entity, as in practice they are controlled by the same organisation.
We also assume that an honest phone requests a TAN at most once.
Dishonest phones may send multiple requests.\end{nnew}

\begin{full}
    \subsection{Example}\label{sec:example}

The following example illustrates our evaluation methodology
from \cref{sec:eval-method} for \dptt{}. Soundness can be encoded
in Tamarin as follows:

\begin{lstlisting}
lemma soundness_v1:
"All idR instClose dayClose tsRisk #tRisk. 
     PClaimAtRisk(idR, dayClose, instClose)@tRisk 
   & Day(tsRisk)@tRisk ==>
// Phone was exposed to a positive patient
// (no misbehaviour)
( Ex idI place dayContag dayTest [#t1..t6].
     IsAt(idR, place, instClose)@t1
   & IsAt(idI, place, instClose)@t2
   & HAClaimInfected(idI, dayContag, dayTest)@t3
   & EarlierDayEq(dayContag, dayClose)@t4
   & EarlierDay(dayClose, dayTest)@t5
   & Within14Days(dayContag, dayTest)@t6
   & not (idR = idI)
   & (All idC cc #t1. PhoneInit(idC, cc)@t1 
          ==> UploadAuth(idC)))
\end{lstlisting}

\begin{full}
  \begin{figure*}%
  \centering
  \includegraphics[width=\textwidth]{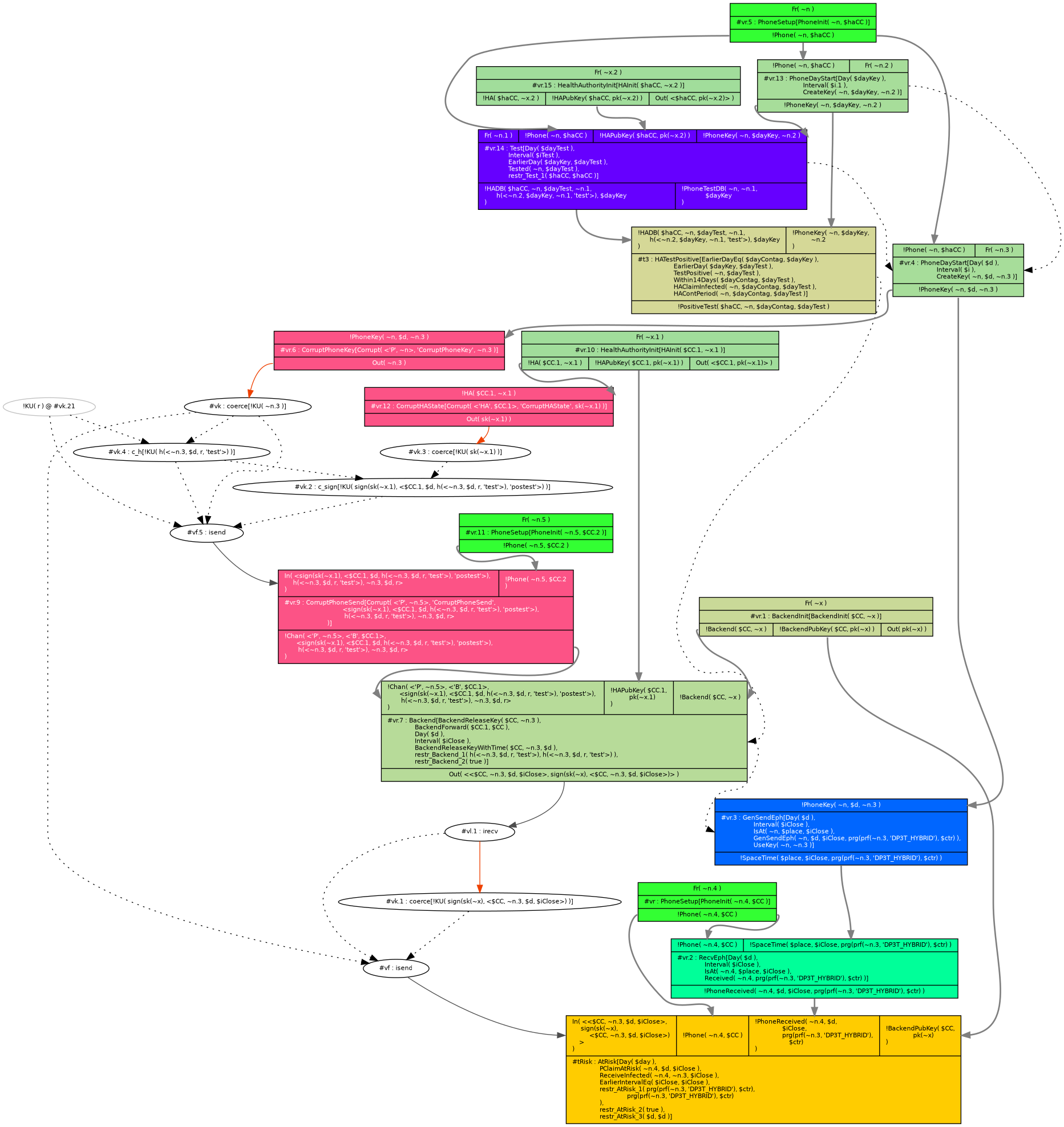}
  \caption{Example trace: Attack against soundness for \dptt{}.}
  \label{fig:example-trace}
  \end{figure*} %
\end{full}

After several minutes of computation, Tamarin returns an attack trace%
\begin{full}
(\cref{fig:example-trace}). To facilitate manual inspection, we
colour-coded each different party and marked the compromise rules in
red. This trace describes a simple attack via malicious upload, i.e. 
\ref{at:dptt-mal-upload}.

\end{full}
We then extract the attack pattern from the trace, which uses the
predicate $\mathit{MaliciusUpload}$ 
to refer to the 
disjunction of all attack patterns against upload authorisation (not
displayed here).
We add the pattern as a disjunction:

\begin{lstlisting}
lemma soundness_v2:
"All idR [..] ==>
// Phone was exposed to a positive patient
// (no misbehaviour)
( Ex idI place dayContag dayTest [#t1..t6].
     IsAt(idR, place, instClose)@t1
   & [..])
| // The key of the infected phone is maliciously 
  // uploaded to the back end.
( Ex idI place dayKey k dayContag dayTest [#t1..t5].
     IsAt(idR, place, instClose)@t1
   & IsAt(idI, place, instClose)@t2
   & HAClaimInfected(idI, dayContag, dayTest)@t3
   & CreateKey(idI, dayKey, k)@t4
   & MaliciousUpload(idI, k)
   & Within14Days(dayContag, dayTest)@t5
   & not (idR = idI)
   & (All idC cc #t1. PhoneInit(idC, cc)@t1 
          ==> UploadAuth(idC)))
\end{lstlisting}

The pattern is formulated to be as descriptive of the attack as
possible, specifying the compromise scenario, but also the relation
between timestamps and other side conditions.
The attack description in \cref{sec:full-attacks} is a textual
representation of the same pattern.
The entry in \cref{tab:attack-surface} describes only the requirements
(compromise steps, time point relations and proximity relations),
whereas some patterns can also detail how parties communicate.

We run Tamarin again and find another attack trace. We inspect this trace
to determine if it should be considered an instance of 
\ref{at:dptt-mal-upload}, which would require generalising the attack
pattern. This is not necessary, and thus we add a pattern to
describe this attack. After finding seven attack patterns, Tamarin
finds no more attacks and the categorisation is complete.

\end{full}

\end{document}
